\documentclass[twocolumn,showpacs,prl,superscriptaddress]{revtex4-1}%
\usepackage{amsfonts}
\usepackage{amsmath}
\usepackage{amssymb}
\usepackage{mathrsfs}
\usepackage{graphicx}
\usepackage{txfonts}
\usepackage{xcolor}%
\usepackage{ulem}
\providecommand{\U}[1]{\protect\rule{.1in}{.1in}}

\begin{document}
\title{Electron-mediated projective quantum nondemolition measurement on a nuclear spin}
\author{Wang Ping}
\affiliation{ College of Education for the future, Beijing Normal University, Zhuhai 519087, China}
\affiliation{Beijing Computational Science Research Center, Beijing 100193, China}
\affiliation{Department of Physics, Centre for Quantum Coherence, and The Hong Kong
Institute of Quantum Information Science and Technology, The Chinese
University of Hong Kong, Shatin, New Territories, Hong Kong, China}
\author{Wen Yang}
\email{wenyang@csrc.ac.cn}
\affiliation{Beijing Computational Science Research Center, Beijing 100193, China}
\author{Renbao Liu}
\email{rbliu@cuhk.edu.hk}
\affiliation{Department of Physics, Centre for Quantum Coherence, and The Hong Kong
Institute of Quantum Information Science and Technology, The Chinese
University of Hong Kong, Shatin, New Territories, Hong Kong, China}

\begin{abstract}
Projective quantum nondemolition (QND) measurement is important for quantum
technologies. Here we propose a method for constructing projective QND
measurement on a nuclear spin via the measurement of an axillary electron spin
in generic electron-nuclear spin systems coupled through weak hyperfine
interaction. The key idea is to apply suitable quantum control on the electron
to construct a weak QND measurement on the nuclear spin and then cascade a
sequence of such measurements into a projective one. We identify a set of
tunable parameters to select the QND observables and control the strength of
the weak QND measurement. We also find that the QND measurement can be
stabilized against realistic experimental control errors. As a demonstration
of our method, we design projective QND measurement on a $^{13}$C nuclear spin
weakly coupled to a nitrogen-vacancy center electron spin in diamond.
\end{abstract}

\maketitle


\section{Introduction}

Measurement backaction \cite{WisemanBook2010} is the measurement-induced
non-unitary disturbance on the system being measured. To overcome the
downgrade of measurement precision caused by the random disturbance, quantum
nondemolition (QND) was introduced
\cite{BraginskiSPU1975,BraginskiiJETP1977,ThornePRL1978,UnruhPRD1979,CavesRMP1980,BraginskyScience1980}%
. The key idea is to measure an observable that is conserved during the free
evolution, so that its value remains unaffected between successive
measurements and sufficient statistics can be build up to improve the
signal-to-noise ratio \cite{ClerkRMP2010}. QND has been demonstrated
experimentally in quantum optics \cite{GrangierNature1998}, microscopic
superconducting systems \cite{LupascuNatPhys2007} and trapped single electron
\cite{PeilPRL1999}.

In hybrid electron-nuclear spin systems, the nuclear spin qubits can be
exploited as quantum resources due to their long coherence times
\cite{LaraouiNatCommun2013,StaudacherNC2015,MaminScience2013,WangNC2017,ZaiserNC2016,ShiNatPhys2014,ShiScience2015,DuNature2009,PfenderNC2017,RosskopfQF2017,SchmittScience2017,GlennNature2018,MuhonenNatNano2014,SaeediScience2013,TyryshkinNatMater2012,PressNatPhys2010}%
. Projective QND measurement on the nuclear spin qubits is important for
scalable quantum computation \cite{TaminiauNatNano2014}, quantum sensing
\cite{PfenderNC2017,RosskopfQF2017} and quantum communication
\cite{YangNatPhoton2016}. The quantum measurement of the nuclear spin is
difficult and is often realized indirectly via measurement of a nearby
electron spin \cite{DreauPRL2013,NeumannScience2010}, e.g., mapping the
population of a \textit{nearby} nuclear spin onto the electron spin via
controlled-NOT gates allows projective QND measurements of the nuclear spin
$z$ component $\hat{I}_{z}$ -- a conserved observable protected by the
external magnetic field \cite{DreauPRL2013,NeumannScience2010}. However,
\textit{remote} nuclear spins weakly coupled to the electron (i.e., coupling
strength $<1/T_{2}^{\ast}$) can hardly be resolved or manipulated to implement
the controlled-NOT gates reliably. Moreover, the transverse components
$(I_{x},I_{y})$ of the nuclear spin are \textit{not} conserved, so their
projective QND measurements are nontrivial. Recently, single-shot readout of
nuclear a spin-1/2 weakly coupled to an nitrogen-vacancy center electron spin
in diamond was demonstrated experimentally by using a series of weak
measurements to trap the target nuclear spin and then read out its state
repeatedly \cite{LiuPRL2017}, but its connection to QND remains unclear.

In this paper, we propose a general method to construct projective QND
measurements on \textit{non-conserved} observables (e.g., the transverse
components)\ of a nuclear spin \textit{weakly} coupled to an axillary electron
spin via hyperfine interaction. The procedure consists of two steps. First,
dynamical decoupling control is applied to the electron spin to establish
electron-nuclear entanglement, followed by a projective measurement on the
electron spin to mediate a single weak measurement on the nuclear spin.
Second, we apply a sequence of such weak measurements to the nuclear spin and
tune the evolution of the nuclear spin between neighboring measurements to
meet the stroboscopic QND condition
\cite{BraginskyJETP1978,ThornePRL1978,CavesRMP1980,BraginskyScience1980,JordanPRB2005,RuskovPRB2005a,JordanPRL2006,AverinPRB2006,JordanPRB2006}%
, so that this sequence of weak QND measurements form a single projective QND
measurement on the nuclear spin. We identify a set of tunable parameters for
flexible, \textit{in situ} control of the QND observable and find optimal
parameters to stabilize the QND measurements against control errors. This work
is relevant to state preparation, quantum sensing, and quantum error
correction via projective QND measurements.

The paper is organized as follows. In Sec. II, we construct a single weak
measurement on the nuclear spin via measurement of the electron spin. In Sec.
III, we use a sequence of weak QND measurements to form a single projective
QND measurement and analyze its stability against control errors. In Sec. IV,
we illustrate our method in a paradigmatic physical system, i.e., a $^{13}$C
nuclear spin weakly coupled to a nitrogen-vacancy center electron spin in
diamond \cite{LiuPRL2017}. In Sec. V, we draw the conclusions.

\section{Electron-mediated measurement on nuclear spin}

\begin{figure}[ptb]
\includegraphics[width=1\columnwidth]{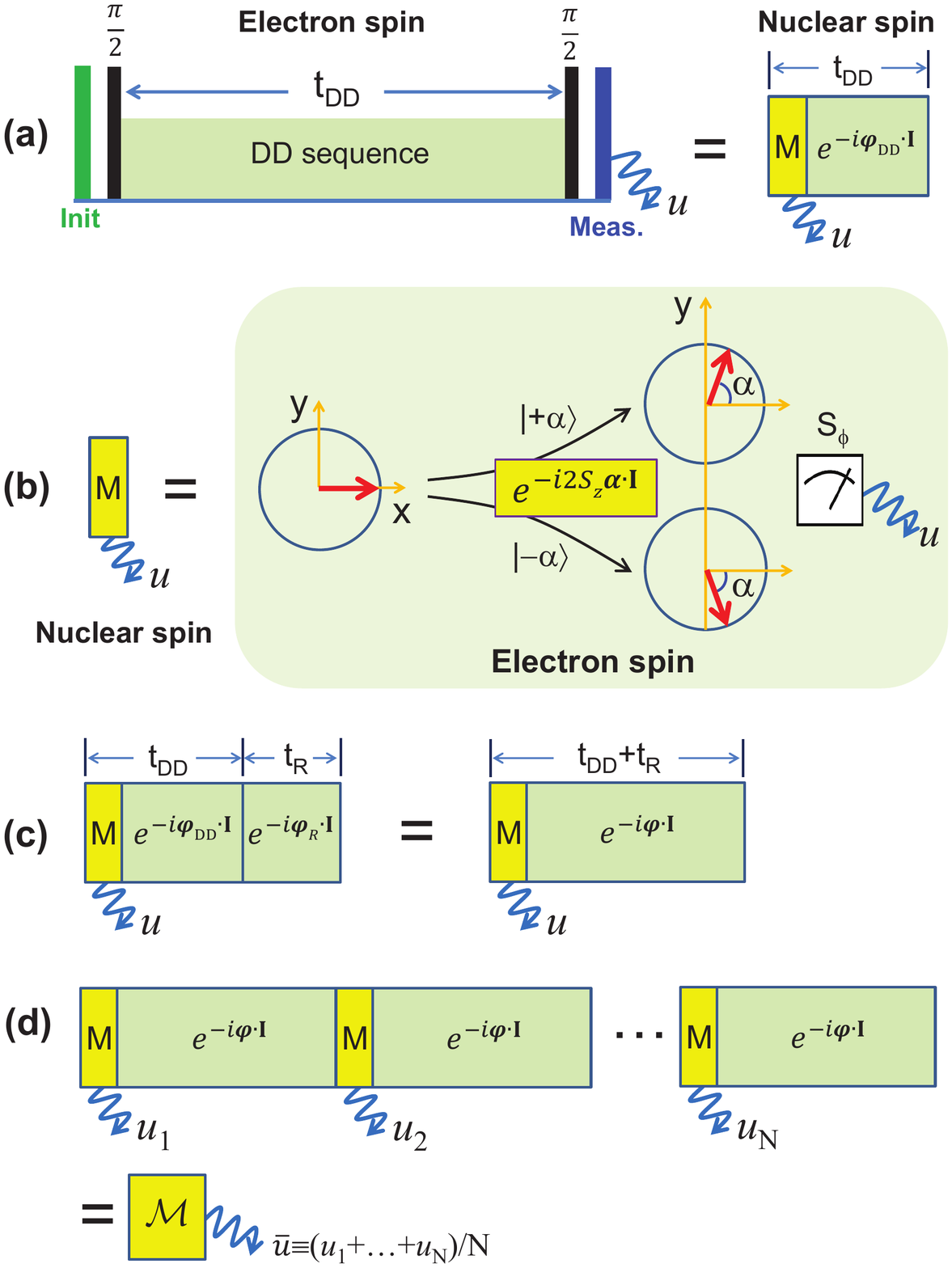}\caption{(a) Control sequence
on an auxillary electron spin coupled to the target nuclear spin via hyperfine
interaction. This mediates an effective measurement (yellow box
\textquotedblleft M\textquotedblright) on the target nuclear spin followed by
a rotation $e^{-i\boldsymbol{\varphi}_{\mathrm{DD}}\cdot\mathbf{\hat{I}}}$.
(b) The effective measurement on the nuclear spin originates from the
conditional rotation of the electron spin (red arrows) around the $z$ axis by
opposite angles for opposite nuclear spin initial states $|\pm\alpha\rangle$
-- the two eigenstates of $\boldsymbol{\hat{\alpha}}\cdot\mathbf{\hat{I}}$,
and the subsequent projective measurement on the electron spin observable
$\hat{S}_{\phi}\equiv\mathbf{\hat{S}}\cdot\mathbf{e}_{\phi}$, where
$\mathbf{e}_{\phi}=\mathbf{e}_{x}\cos\phi+\mathbf{e}_{y}\sin\phi$. (c)
Immediately after the $\hat{S}_{\phi}$ measurement, we re-initialize the
electron spin into $|+\rangle$ and append a waiting time $t_{R}$, during which
we flip the electron spin to engineer the nuclear spin evolution
$e^{-i\boldsymbol{\varphi}_{R}\cdot\mathbf{\hat{I}}}$. The total evolution of
the nuclear spin after the measurement becomes $e^{-i\boldsymbol{\varphi}%
\cdot\mathbf{\hat{I}}}\equiv e^{-i\boldsymbol{\varphi}_{R}\cdot\mathbf{\hat
{I}}}e^{-i\boldsymbol{\varphi}_{\mathrm{DD}}\cdot\mathbf{\hat{I}}}$. (d)
Cascading $N$ weak measurements with outcome $(u_{1},\cdots,u_{N})$ into a
single strong measurement with outcome $\bar{u}\equiv(u_{1}+\cdots+u_{N})/N$.}%
\label{G_SETUP}%
\end{figure}

We consider a target nuclear spin-1/2 $\hat{\mathbf{I}}$ with Zeeman
Hamiltonian $\omega_{n}\hat{I}_{z}$ coupled to an auxillary electron spin with
Hamiltonian $\omega_{0}\hat{S}_{z}$ through the hyperfine interaction. The
hyperfine interaction is usually much weaker than $\left\vert \omega
_{0}\right\vert $, so it does not cause electron spin flip and hence can be
written as $|+_{z}\rangle\langle+_{z}|\mathbf{a}_{+}\cdot\hat{\mathbf{I}%
}+|-_{z}\rangle\langle-_{z}|\mathbf{a}_{-}\cdot\hat{\mathbf{I}}$, where
$|\pm_{z}\rangle$ are the two eigenstates of the electron spin $\hat{S}_{z}$.
In the interaction picture of the electron spin, the total Hamiltonian takes
the form%

\begin{equation}
\hat{H}=\boldsymbol{\omega}\cdot\hat{\mathbf{I}}+\hat{S}_{z}\mathbf{A}%
\cdot\hat{\mathbf{I}}\mathbf{,} \label{HAMIL}%
\end{equation}
with $\boldsymbol{\omega}\equiv\omega_{n}\mathbf{e}_{z}+(\mathbf{a}%
_{+}+\mathbf{a}_{-})/2$ the hyperfine-shifted nuclear Larmor frequency and
$\mathbf{A}\equiv\mathbf{a}_{+}-\mathbf{a}_{-}$ the effective hyperfine interaction.

\subsection{Theoretical formalism}

\label{SEC_PHYSICS}

The protocol for constructing a single electron-mediated weak measurement on
the nuclear spin is shown in Fig. \ref{G_SETUP}(a). At $t=0$, the electron
spin starts from the spin-up state $|+_{z}\rangle$ along the $+z$ axis, while
the nuclear spin starts from a general initial state $\hat{\rho}$. Next, a
$\pi/2$-pulse around the $y$ axis is applied to the electron spin to rotate it
to the $\hat{S}_{x}=+1/2$ eigenstate $|+_{x}\rangle$. Next a dynamical
decoupling (DD) sequence is applied to the electron spin to generate the
evolution
\begin{equation}
\hat{U}_{\mathrm{DD}}=e^{-i\boldsymbol{\varphi}_{\mathrm{DD}}\cdot
\hat{\mathbf{I}}}e^{-i2\hat{S}_{z}\boldsymbol{\alpha}\cdot\hat{\mathbf{I}}},
\label{UDD}%
\end{equation}
driving the electron-nuclear system into the entangled state $\hat
{U}_{\mathrm{DD}}|+_{x}\rangle\hat{\rho}\langle+_{x}|\hat{U}_{\mathrm{DD}%
}^{\dagger}$. Here $\boldsymbol{\varphi}_{\mathrm{DD}}$ and
$\boldsymbol{\alpha}$ are vectors that depend on the hyperfine interaction
$\mathbf{A}$ and the DD sequence (see Appendix A). Finally a projective
measurement is made on the electron spin $\hat{S}_{\phi}\equiv\mathbf{\hat
{S}\cdot e}_{\phi}$ along $\mathbf{e}_{\phi}\equiv\mathbf{e}_{x}\cos
\phi+\mathbf{e}_{y}\sin\phi$ by first applying a $\pi/2$-pulse along
$\mathbf{e}_{\phi-\pi/2}$ to rotate $\hat{S}_{\phi}$ to $\hat{S}_{z}$ and then
measuring $\hat{S}_{z}$. Upon getting a specific outcome $u$ ($=+$ or $-$),
the \textit{nuclear spin} collapses into the $u$-dependent (un-normalized)
final state
\begin{equation}
\langle u_{\phi}|\hat{U}_{\mathrm{DD}}|+_{x}\rangle\hat{\rho}\langle+_{x}%
|\hat{U}_{\mathrm{DD}}^{\dagger}|u_{\phi}\rangle=e^{-i\boldsymbol{\varphi
}_{\mathrm{DD}}\cdot\hat{\mathbf{I}}}\hat{M}_{u}\hat{\rho}(\hat{M}%
_{u})^{\dagger}e^{i\boldsymbol{\varphi}_{\mathrm{DD}}\cdot\hat{\mathbf{I}}},
\label{RHO_U}%
\end{equation}
where $|\pm_{\phi}\rangle$ are the two eigenstates of the electron spin
$\hat{S}_{\phi}$ and
\begin{equation}
\hat{M}_{u}\equiv\langle u_{\phi}|e^{-i2\hat{S}_{z}\boldsymbol{\alpha}%
\cdot\hat{\mathbf{I}}}|+_{x}\rangle=\frac{e^{i(\boldsymbol{\alpha}\cdot
\hat{\mathbf{I}}-\phi/2)}+ue^{-i(\boldsymbol{\alpha}\cdot\hat{\mathbf{I}}%
-\phi/2)}}{2} \label{MU}%
\end{equation}
is the $u$-dependent Kraus operator acting on the \textit{nuclear spin}. When
$\phi=0$, $\{\hat{M}_{u}\}$ reduce to those in Ref. \cite{GreinerSciRep2017}.

The nuclear spin evolution from $\hat{\rho}$ at $t=0$ to the $\hat{u}%
$-dependent final state [Eq. (\ref{RHO_U})] consists of a measurement followed
by a unitary rotation $e^{-i\boldsymbol{\varphi}_{\mathrm{DD}}\cdot
\hat{\mathbf{I}}}$ [see Fig. \ref{G_SETUP}(a)]. The measurement is described
by the positive-operator valued measure (POVM) set $\{\hat{M}_{\pm}\}$, i.e.,
outcome $u$ occurs with probability
\[
P(u)=\operatorname*{Tr}\hat{M}_{u}\hat{\rho}\hat{M}_{u}^{\dagger}%
=P(u|\alpha)\langle\alpha|\hat{\rho}|\alpha\rangle+P(u|-\alpha)\langle
-\alpha|\hat{\rho}|-\alpha\rangle
\]
and collapses the nuclear spin into $\hat{M}_{u}\hat{\rho}\hat{M}_{u}%
^{\dagger}$, where
\begin{equation}
P(u|\pm\alpha)=\frac{1+u\cos(\phi\mp\alpha)}{2} \label{PUA}%
\end{equation}
is the conditional probability of outcome $u$ for the nuclear spin initial
state $|\pm\alpha\rangle$. In the above, we have defined $\alpha
\equiv|\boldsymbol{\alpha}|$, $\hat{\boldsymbol{\alpha}}\equiv
\boldsymbol{\alpha}/\alpha$, and used $|\pm\alpha\rangle$ for the eigenstates
of $\hat{\boldsymbol{\alpha}}\cdot\hat{\mathbf{I}}$ with eigenvalue $\pm1/2$.

\subsection{Physical picture}

The measurement on the target nuclear spin is constructed by first entangling
it with an ancillary electron spin via the DD-generated evolution
$e^{-i2\hat{S}_{z}\boldsymbol{\alpha}\cdot\hat{\mathbf{I}}}$ and then
measuring this electron spin. As shown in Fig. \ref{G_SETUP}(b), the evolution
$e^{-i2\hat{S}_{z}\boldsymbol{\alpha}\cdot\hat{\mathbf{I}}}$ is a
nuclear-controlled rotation of the electron spin around the $z$ axis: for
opposite nuclear spin initial states $|\pm\alpha\rangle$, it rotates the
electron spin from $|+_{x}\rangle$ to\ $|\pm\alpha\rangle_{\mathrm{e}}\equiv
e^{\mp i\alpha\hat{S}_{z}}|+_{x}\rangle$. This correlates the two orthogonal
eigenstates $|\pm\alpha\rangle$ of the nuclear spin observable $\hat
{\boldsymbol{\alpha}}\cdot\hat{\mathbf{I}}$ with \textit{distinct} (but not
necessarily orthogonal) final states $|\pm\alpha\rangle_{\mathrm{e}}$ of the
electron spin. In the subsequent measurement on the electron spin, $|\pm
\alpha\rangle_{\mathrm{e}}$ in turn give \textit{distinct} measurement
distributions $P(u|\pm\alpha)$, i.e., measuring the electron spin can
distinguish (partially) between $|\pm\alpha\rangle_{\mathrm{e}}$ or
equivalently between $|\pm\alpha\rangle$. So measuring the electron spin
observable $\hat{S}_{\phi}$ mediates a (partial) measurement on the nuclear
spin observable $\hat{\boldsymbol{\alpha}}\cdot\hat{\mathbf{I}}$. The strength
of the measurement on the nuclear spin is quantified by the degree to which
this measurement can distinguish between the two eigenstates $|\pm
\alpha\rangle$ of $\hat{\boldsymbol{\alpha}}\cdot\hat{\mathbf{I}}$, i.e., the
distinguishability between $P(u|\pm\alpha)$.

Quantitatively, we characterize $P(u|a)$ ($a=\pm\alpha$) by the conditional
expectation value of the measurement outcome
\begin{equation}
\langle u\rangle_{a}\equiv P(+|a)-P(-|a)=\cos(\phi-a) \label{UA}%
\end{equation}
and its fluctuation%
\begin{equation}
\sigma_{a}\equiv\sqrt{\langle u^{2}\rangle_{a}-\langle u\rangle_{a}^{2}}%
=|\sin(\phi-a)|. \label{SA}%
\end{equation}
We identify $\langle u\rangle_{\alpha}-\langle u\rangle_{-\alpha}$ and
$\sigma_{\alpha}+\sigma_{-\alpha}$, respectively, as the \textquotedblleft
signal\textquotedblright\ and \textquotedblleft noise\textquotedblright\ for
distinguishing between the two eigenstates $|\pm\alpha\rangle$ of
$\hat{\boldsymbol{\alpha}}\cdot\hat{\mathbf{I}}$ and define the
distinguishability between $P(u|\pm\alpha)$ or equivalently the strength of
the measurement on $\hat{\boldsymbol{\alpha}}\cdot\hat{\mathbf{I}}$ as the
signal-to-noise ratio:%
\begin{equation}
D\equiv\frac{\langle u\rangle_{\alpha}-\langle u\rangle_{-\alpha}}%
{\sigma_{\alpha}+\sigma_{-\alpha}}=\min\{|\tan\alpha|,|\tan\phi|\}. \label{D}%
\end{equation}
Here $\tan\alpha$ characterizes the difference bewteen $|\pm\alpha
\rangle_{\mathrm{e}}$ or equivalently the degree of electron-nuclear
entanglement, while $\tan\phi$ characterizes the degree to which measuring the
electron spin observable $\hat{S}_{\phi}$ can distinguish between $|\pm
\alpha\rangle_{\mathrm{e}}$ and hence $|\pm\alpha\rangle$. For example,
$\tan\alpha\approx0$ leads to $|+\alpha\rangle_{\mathrm{e}}\approx
|-\alpha\rangle_{\mathrm{e}}$ up to a phase factor, so any measurement on the
electron spin always yields $P(u|+\alpha)\approx P(u|-\alpha)$ and $D\approx
0$, i.e., weak measurement on the nuclear spin. For a given $\alpha$,
measuring $\hat{S}_{y}$ (i.e., $\phi=\pi/2$) can optimally distinguish
$|\pm\alpha\rangle_{\mathrm{e}}$ by yielding maximally different
$P(u|\pm\alpha)$, so it gives maximal measurement strength. By contrast,
measuring $\hat{S}_{x}$ (i.e., $\phi=0$) cannot distinguish $|\pm\alpha
\rangle_{\mathrm{e}}$ since it always gives $P(u|\alpha)=P(u|-\alpha)$, so it
leads to vanishing measurement strength: $D=0$.

In the above, we have assumed ideal projective measurement on the electron
spin observable $\hat{S}_{\phi}$. If this measurement is not perfect, as
quantified by a finite probability $p_{\pm}$ to get an outcome $u=\pm$ when
the true electron spin state is $|\pm_{\phi}\rangle$, then we should replace
$\cos(\phi\mp\alpha)$ in Eq. (\ref{PUA}) by $\Delta p+\bar{p}\cos(\phi
\mp\alpha)$, where $\Delta p\equiv p_{+}-p_{-}$ and $\bar{p}\equiv p_{+}%
+p_{-}-1$. This reduces the \textquotedblleft signal\textquotedblright\ by a
factor $\bar{p}$, but increases the \textquotedblleft noise\textquotedblright,
so it weakens the measurement strength to%
\begin{equation}
D=\frac{2\bar{p}|\sin\alpha\sin\phi|}{%
{\displaystyle\sum\limits_{a=\pm\alpha}}
\sqrt{1-\left[  \Delta p+\bar{p}\cos(\phi-a)\right]  ^{2}}}\overset{\bar{p}%
\ll1}{\approx}\bar{p}|\sin\alpha\sin\phi|. \label{DLOW}%
\end{equation}

To summarize, the measurement on the nuclear spin is controlled by two
parameters $\boldsymbol{\alpha}=\alpha\boldsymbol{\hat{\alpha}}$ and $\phi$,
i.e., $\boldsymbol{\hat{\alpha}}$ controls the nuclear spin observable
$\hat{\boldsymbol{\alpha}}\cdot\hat{\mathbf{I}}$ being measured, while
$\tan\alpha$ and $\tan\phi$ controls the distinguishability $D$ between
$P(u|\pm\alpha)$ or equivalently the measurement strength. For example,
$\tan\alpha\tan\phi\approx0$ gives nearly identical $P(u|\pm\alpha)$ and hence
$D\approx0$, while $\alpha=\phi=\pi/2$ gives \textit{non-overlapping }(i.e.,
perfectly distinguishable) $P(u|\alpha)=\delta_{u,+}$ and $P(u|-\alpha
)=\delta_{u,-}$ and hence a projective measurement ($D=\infty$), as described
by the POVM set [Eq. (\ref{MU})] $\hat{M}_{+}\equiv|+\alpha\rangle
\langle+\alpha|$ and $\hat{M}_{-}\equiv|-\alpha\rangle\langle-\alpha|$. The
parameter $\phi$ can be tuned directly in the experiment, while the direction
and magnitude of $\boldsymbol{\alpha}$ can be tuned \textit{independently} by
varying the duration and structure of the DD sequence. Next we discuss this
tunability in more detail.

\subsection{Tunability of $\boldsymbol{\alpha}$}

The tunability of $\boldsymbol{\alpha}$ becomes physically transparent when
the perpendicular part $\mathbf{A}_{\perp}$ of the hyperfine interaction
vector $\mathbf{A}$ with respect to $\boldsymbol{\omega}$ is much smaller than
$|\boldsymbol{\omega}|$. In this case, we obtain approximate analytical
expressions (see Appendix B)
\begin{align}
\boldsymbol{\varphi}_{\mathrm{DD}}  &  \approx\boldsymbol{\omega
}t_{\mathrm{DD}},\label{PHI_P}\\
\boldsymbol{\alpha}  &  \approx|f_{\mathrm{DD}}|\boldsymbol{\mathbb{R}}%
(-\arg(f_{\mathrm{DD}})\frac{\boldsymbol{\omega}}{|\boldsymbol{\omega}|}%
)\frac{\mathbf{A}_{\perp}t_{\mathrm{DD}}}{2}, \label{ALPHA}%
\end{align}
where $\boldsymbol{\mathbb{R}}(\boldsymbol{\theta})$ is the SO(3) rotation
matrix that rotates a vector around the axis $\boldsymbol{\theta}$ by an angle
$|\boldsymbol{\theta}|$ and%
\begin{equation}
f_{\mathrm{DD}}\equiv\frac{1}{t_{\mathrm{DD}}}\int_{0}^{t_{\mathrm{DD}}%
}s(t)e^{i|\boldsymbol{\omega}|t}dt \label{FDD}%
\end{equation}
accounts for the DD sequence \cite{CywinskiPRB2008,YangRPP2017}:\ $s(t)$
starts from $s(0)=+1$ and switches its sign at the timings of each $\pi$-pulse
in the DD sequence. The tunability of $\boldsymbol{\alpha}$ is completely
characterized by $f_{\mathrm{DD}}$: its phase $\arg(f_{\mathrm{DD}})$ controls
the direction of $\boldsymbol{\alpha}$ and hence the nuclear spin observable
$\hat{\boldsymbol{\alpha}}\cdot\mathbf{\hat{I}}$ to be measured, while its
magnitude $|f_{\mathrm{DD}}|$ controls the magnitude of $\boldsymbol{\alpha}$
and hence the measurement strength. By varying the duration of the DD sequence
and the timings of the constituent $\pi$-pulses, we can tune $\arg
(f_{\mathrm{DD}})$ and $|f_{\mathrm{DD}}|$ independently, e.g., the
$N_{\mathrm{DD}}$-period Carr--Purcell--Meiboom--Gill (CPMG) sequence
$(\tau/4$-$\pi$-$\tau/2$-$\pi$-$\tau/4)^{N_{\mathrm{DD}}}$ corresponds to
$t_{\mathrm{DD}}\equiv N_{\mathrm{DD}}\tau$ and
\[
f_{\mathrm{DD}}=-e^{i|\boldsymbol{\omega}|t_{\mathrm{DD}}/2}\frac
{4}{|\boldsymbol{\omega}|t_{\mathrm{DD}}}\frac{\sin^{2}(|\boldsymbol{\omega
}|\tau/8)\sin(|\boldsymbol{\omega}|t_{\mathrm{DD}}/2)}{\cos
(|\boldsymbol{\omega}|\tau/4)},
\]
so $\arg(f_{\mathrm{DD}})$ can be tuned by varying $N_{\mathrm{DD}}$, while
$|f_{\mathrm{DD}}|$ can be tuned by varying $\tau$. Setting $\tau
=2\pi/|\boldsymbol{\omega}|$ gives maximal $f_{\mathrm{DD}}=2/\pi$ and hence
maximal $\boldsymbol{\alpha}=\mathbf{A}_{\perp}t_{\mathrm{DD}}/\pi$.

Equation (\ref{ALPHA}) suggests that the magnitude of $\boldsymbol{\alpha}$
can be enhanced indefinitely by increasing the total duration $t_{\mathrm{DD}%
}$ of the DD sequence. In practice, however, the ancillary electron spin,
albeit under the DD control, still has a finite coherence time $T_{2}$, which
sets an upper limit $t_{\mathrm{DD}}\lesssim T_{2}$ and hence $\alpha
\lesssim|\mathbf{A}_{\perp}|T_{2}/2$ since $|f_{\mathrm{DD}}|\leq1$. Here, in
addition to providing the desired tunability for $\boldsymbol{\alpha}$, the DD
sequence also filters out undesirable noises to enhance the electron spin
coherence time $T_{2}$ beyond the inhomogeneous dephasing time $T_{2}^{\ast}$.
This not only provides a better spectral resolution $1/T_{2}$ ($\ll
1/T_{2}^{\ast}$) to single out the target nuclear spin among other
environmental nuclei \cite{LiuPRL2017}, but also enhances the maximal
achievable measurement strength. As a result, projective measurement (i.e.,
$\alpha=\phi=\pi/2$) can be achieved even for weakly coupled target nuclear
spins with hyperfine interaction $|\mathbf{A}|<1/T_{2}^{\ast}$, as long as
$|\mathbf{A}_{\perp}|\gtrsim1/T_{2}$. For extremely weak hyperfine interaction
$|\mathbf{A}_{\perp}|\lesssim1/T_{2}$, the maximal achievable $\alpha$ is less
than $\pi/2$, i.e., a single projective measurement on the electron spin can
at most mediate a weak measurement on the nuclear spin. Interestingly, even in
this case, projective measurement is still possible.

\section{Construction of projective QND measurements}

When only weak measurements are available, a natural idea to construct a
projective measurement is to cascade a sequence of weak measurements into a
single projective measurement \cite{ClerkRMP2010,LiuAdvPhys2010}: although
each measurement is weak, the two eigenstates $|\pm\alpha\rangle$ of the
observable $\boldsymbol{\hat{\alpha}}\cdot\mathbf{\hat{I}}$ being measured can
always be distinguished reliably by the statistics of a large number of
repeated weak measurements on $|\pm\alpha\rangle$. In our case, however, the
weak measurement over $\boldsymbol{\hat{\alpha}}\cdot\mathbf{\hat{I}}$ is
always followed by a unitary rotation $e^{-i\boldsymbol{\varphi}_{\mathrm{DD}%
}\cdot\mathbf{\hat{I}}}$. Since $\boldsymbol{\varphi}_{\mathrm{DD}}$ and
$\boldsymbol{\hat{\alpha}}$ are usually noncollinear, a naive repetition of
the protocol in Fig. \ref{G_SETUP}(a) would not form a projective measurement,
because $e^{-i\boldsymbol{\varphi}_{\mathrm{DD}}\cdot\mathbf{\hat{I}}}$ may
destroy the eigenstates $|\pm\alpha\rangle$ of the observable
$\boldsymbol{\hat{\alpha}}\cdot\mathbf{\hat{I}}$ and hence make repeated
measurements on $|\pm\alpha\rangle$ impossible. To construct a projective
measurement from a sequence of weak measurements, the first step is to protect
the eigenstates of $\boldsymbol{\hat{\alpha}}\cdot\mathbf{\hat{I}}$ against
the rotation $e^{-i\boldsymbol{\varphi}_{\mathrm{DD}}\cdot\mathbf{\hat{I}}}$,
i.e., to make each weak measurement QND.

\subsection{Stroboscopic QND condition}

The idea of stroboscopic QND
\cite{BraginskyJETP1978,ThornePRL1978,CavesRMP1980,BraginskyScience1980,JordanPRB2005,RuskovPRB2005a,JordanPRL2006,AverinPRB2006,JordanPRB2006}
is to use \textit{stroboscopic} measurements with precise timings to protect
the eigenstates of the observable being measured. For example, under an
external magnetic field $B$ along the $z$ axis, the Zeeman Hamiltonian
$\hat{H}_{0}=\gamma B\hat{S}_{z}$ of a spin-1/2 drives periodic Larmor
precession, so the evolution operator $e^{-i\gamma B\tau\hat{S}_{z}}$ becomes
a c-number when the evolution time $\tau$ equals the Larmor precession period
$2\pi/(\gamma B)$. If we perform a sequence of stroboscopic measurements with
the measurement interval being an integer multiple of the Larmor precession
period \cite{JordanPRB2005,JordanPRL2006,AverinPRB2006,JordanPRB2006}, then
the evolution between neighboring measurements does not affect the
post-measurement state, so the QND condition is satisfied.

Interestingly, although our protocol in Fig. \ref{G_SETUP}(a) is not
stroboscopic, it effectively generates a \textquotedblleft
stroboscopic\textquotedblright\ measurement on the nuclear spin observable
$\boldsymbol{\hat{\alpha}}\cdot\mathbf{\hat{I}}$, but is followed by a unitary
rotation $e^{-i\boldsymbol{\varphi}_{\mathrm{DD}}\cdot\mathbf{\hat{I}}}$ that
may destroy the eigenstates $|\pm\alpha\rangle$ of $\boldsymbol{\hat{\alpha}%
}\cdot\mathbf{\hat{I}}$. To protect $|\pm\alpha\rangle$, we re-initialize the
electron spin into $|+_{z}\rangle$ immediately after the $\hat{S}_{\phi}$
measurement and then append a waiting time $t_{R}$ [see Fig. \ref{G_SETUP}%
(c)]. Since the nuclear spin precession frequency is $\boldsymbol{\omega}%
\pm\mathbf{A}/2$ when the electron stays in $|\pm_{z}\rangle$, we can engineer
the nuclear spin evolution $e^{-i\boldsymbol{\varphi}_{R}\cdot\mathbf{\hat{I}%
}}$ during this waiting time by flipping the electron spin between $|\pm
_{z}\rangle$ with $\pi$-pulses, e.g., $e^{-i\boldsymbol{\varphi}_{R}%
\cdot\mathbf{\hat{I}}}=e^{-i(\boldsymbol{\omega}+\mathbf{A}/2)t_{R}%
\cdot\mathbf{\hat{I}}}$ if we do not flip the electron spin and
$e^{-i\boldsymbol{\varphi}_{R}\cdot\mathbf{\hat{I}}}=e^{-i(\boldsymbol{\omega
}-\mathbf{A}/2)(t_{R}-t_{1})\cdot\mathbf{\hat{I}}}e^{-i(\boldsymbol{\omega
}+\mathbf{A}/2)t_{1}\cdot\mathbf{\hat{I}}}$ if we flip the electron spin at
$t_{1}$. Since the two rotation axes $\boldsymbol{\omega}\pm\mathbf{A}/2$ are
usually non-collinear, we can achieve an \textit{arbitrary} evolution
$e^{-i\boldsymbol{\varphi}_{R}\cdot\mathbf{\hat{I}}}$ by tuning $t_{R}$ and
the timings of the electron spin flip. The total evolution of the nuclear spin
after each weak measurement becomes%
\begin{equation}
e^{-i\boldsymbol{\varphi}\cdot\mathbf{\hat{I}}}\equiv e^{-i\boldsymbol{\varphi
}_{R}\cdot\mathbf{\hat{I}}}e^{-i\boldsymbol{\varphi}_{\mathrm{DD}}%
\cdot\mathbf{\hat{I}}}. \label{PHI_DEF}%
\end{equation}
Protecting the eigenstates $|\pm\alpha\rangle$ of $\boldsymbol{\hat{\alpha}%
}\cdot\mathbf{\hat{I}}$ requires $e^{-i\boldsymbol{\varphi}\cdot
\mathbf{\hat{I}}}$ to commute with $\boldsymbol{\hat{\alpha}}\cdot
\mathbf{\hat{I}}$, or equivalently%
\begin{equation}
\operatorname{mod}(|\boldsymbol{\varphi}|,2\pi)=0\ \ \ \ \ \ \mathrm{or}%
\ \ \ \ \ \ \ \boldsymbol{\varphi}\parallel\boldsymbol{\hat{\alpha}}.
\label{QND}%
\end{equation}
For weak hyperfine interaction and hence $\boldsymbol{\varphi}\perp
\boldsymbol{\hat{\alpha}}$, Eq. (\ref{QND}) reduces to $\operatorname{mod}%
(|\boldsymbol{\varphi}|,2\pi)=0$, reminiscent of stroboscopic QND measurements
at integer multiples of the system's period
\cite{JordanPRB2005,JordanPRL2006,AverinPRB2006,JordanPRB2006}. In Eq.
(\ref{PHI_DEF}), $e^{-i\boldsymbol{\varphi}_{\mathrm{DD}}\cdot\mathbf{\hat{I}%
}}$ is determined by the DD sequence [Fig. \ref{G_SETUP}(a)], so the QND
condition Eq. (\ref{QND}) imposes different requirements on
$e^{-i\boldsymbol{\varphi}_{R}\cdot\mathbf{\hat{I}}}$ for different DD
sequences. Due to the complete tunability in $e^{-i\boldsymbol{\varphi}%
_{R}\cdot\mathbf{\hat{I}}}$, the QND condition is always achievable. Moreover,
as we prove in Appendix \ref{APPEND_PHIR}, for a large class of DD sequences,
i.e., when the DD sequence is the repetition of an even-order concatenated DD
\cite{KhodjastehPRL2005,YaoPRL2007,KhodjastehPRA2007,YangFP2011} (with the
widely-used CPMG sequence $(\tau/4$-$\pi$-$\tau/2$-$\pi$-$\tau
/4)^{N_{\mathrm{DD}}}$ being an example), the QND condition can be achieved by
tuning $t_{R}$ only, without flipping the electron spin during the waiting time.

Under the QND condition, we can repeat the structure in Fig. \ref{G_SETUP}(c)
to form a single projective measurement, as shown in Fig. \ref{G_SETUP}(d). In
the following, we first describe the gradual formation of a single projective
QND measurement from a sequence of weak QND measurements
\cite{JordanPRB2006,LiuAdvPhys2010} and then discuss its stability against
control errors.

\subsection{Cascading weak measurements into projective measurement}

To begin with, we use the eigenstates $|\pm\alpha\rangle$ of the observable
$\boldsymbol{\hat{\alpha}}\cdot\mathbf{\hat{I}}$ to rewrite the POVM\ elements
$\{\hat{M}_{u}\}$ [Eq. (\ref{MU})] as%
\[
\hat{M}_{u}\equiv\sum_{a=\pm\alpha}\sqrt{P(u|a)}e^{i\theta_{u,a}}%
|a\rangle\langle a|,
\]
where $e^{i\theta_{u,a}}$ is a trivial phase factor and $P(u|a)$ is the
conditional probability of outcome $u$ for the initial state $|a\rangle$
($a=\pm\alpha$), see Eq. (\ref{PUA}). As discussed at the end of Sec.
\ref{SEC_PHYSICS}, weak measurement ($D\ll1$) corresponds to small difference
between $P(u|\pm\alpha)$, so that a single measurement can barely distinguish
between $|\pm\alpha\rangle$; while a projective measurement ($D\rightarrow
\infty$) corresponds to non-overlapping $P(u|\pm\alpha)$, so that a single
measurement can perfectly distinguish between $|\pm\alpha\rangle$.

\begin{figure}[ptb]
\includegraphics[width=\columnwidth]{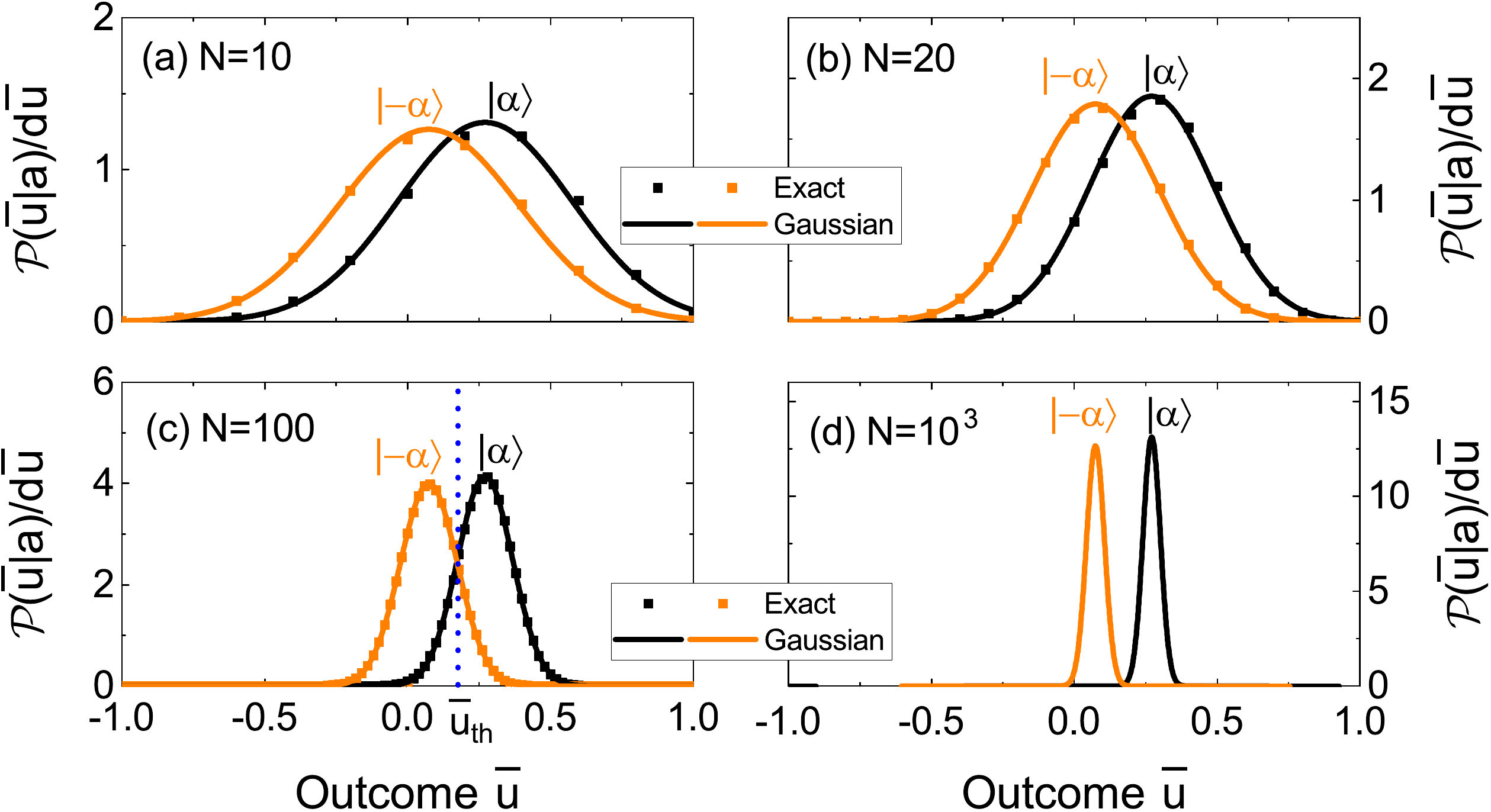}\caption{Conditional
distributions $\mathcal{P}(\bar{u}|\pm\alpha)$ for the average $\bar
{u}=(1/N)\sum_{i=1}^{N}u_{i}$ of $N$ sequential binary measurements with
$\alpha=0.1$ and $\phi=4\pi/9$, i.e., the strength of each binary measurement
is $D\approx0.1$. Squares for Eq. (\ref{P_EXACT}) and solid lines for Eq.
(\ref{P_GAUSSIAN}). The vertical dashed line in (c) marks the optimal
threshold $\bar{u}_{\mathrm{th}}$ that maximizes the average readout
fidelity.}%
\label{G_DISTRIBUTION}%
\end{figure}

As shown in Fig. \ref{G_SETUP}(d), we consider a squence of $N$ identical
measurements with an outcome $\mathbf{u}\equiv(u_{1},\cdots,u_{N})$, where
$u_{i}$ is the outcome of the $i$th measurement. To keep our theory general,
we do not make any assumptions about the strength of each measurement. The
POVM element for these $N$ measurements is $\hat{M}_{\mathbf{u}}\equiv
e^{-i\boldsymbol{\varphi}\cdot\mathbf{\hat{I}}}\hat{M}_{u_{N}}\cdots
e^{-i\boldsymbol{\varphi}\cdot\mathbf{\hat{I}}}\hat{M}_{u_{1}}$, e.g., for an
arbitrary initial state $\hat{\rho}$, the probability for outcome $\mathbf{u}$
is $\operatorname*{Tr}\hat{M}_{\mathbf{u}}\hat{\rho}\hat{M}_{\mathbf{u}%
}^{\dagger}$. Under the QND condition Eq. (\ref{QND}), $\hat{M}_{+},\hat
{M}_{-}$, and $e^{-i\boldsymbol{\varphi}\cdot\mathbf{\hat{I}}}$ mutually
commute, so $\hat{M}_{\mathbf{u}}=e^{-iN\boldsymbol{\varphi}\cdot
\mathbf{\hat{I}}}\hat{M}_{+}^{N_{+}}\hat{M}_{-}^{N_{-}}$ only depends on the
number $N_{\pm}$ of outcome $\pm$ contained in $\mathbf{u}$, or equivalently
the averaged outcome
\[
\bar{u}\equiv\frac{N_{+}-N_{-}}{N}=\frac{1}{N}\sum_{i=1}^{N}u_{i},
\]
which takes discrete values in $[-1,+1]$ evenly spaced by $d\bar{u}\equiv2/N$.
This allows us to regard the $N$ sequential measurements with outcome
$\mathbf{u}$ as a \textit{single} effective measurement with outcome $\bar{u}$
[Fig. \ref{G_SETUP}(d)], without losing any information about the
post-measurement state. Since each constituent measurement gives a
binary-valued outcome $u=+$ or $-$, while the resulting effective measurement
gives a multi-valued outcome $\bar{u}$, we call the former \textit{binary}
measurement and the latter \textit{multi-outcome} measurement to avoid
confusion. Since there are $N!/(N_{+}!N_{-}!)$ distinct $\mathbf{u}$'s that
give the same $\bar{u}$, the POVM element for the multi-outcome measurement
is
\[
\mathcal{\hat{M}}_{\bar{u}}\equiv\sqrt{\frac{N!}{N_{+}!N_{-}!}}\hat
{M}_{\mathbf{u}}=\sum_{a=\pm\alpha}\sqrt{\mathcal{P}(\bar{u}|a)}%
e^{i\Theta_{\bar{u},a}}|a\rangle\langle a|,
\]
where $e^{i\Theta_{\bar{u},a}}$ is a trivial phase factor and
\begin{equation}
\mathcal{P}(\bar{u}|a)\equiv N!\frac{[P(+|a)]^{N_{+}}}{N_{+}!}\frac
{[P(-|a)]^{N_{-}}}{N_{-}!} \label{P_EXACT}%
\end{equation}
is the probability of outcome $\bar{u}$ conditioned on the initial state being
$|a\rangle$ ($a=\pm\alpha)$. When $N\gg1$, the conditional distribution of
$\bar{u}$ becomes Gaussian (see Fig. \ref{G_DISTRIBUTION}):
\begin{equation}
\mathcal{P}(\bar{u}|a)\approx\frac{e^{-(\bar{u}-\langle u\rangle_{a}%
)^{2}/(2\sigma_{a}^{2}/N)}}{\sqrt{2\pi}\sigma_{a}/\sqrt{N}}d\bar{u}.
\label{P_GAUSSIAN}%
\end{equation}
According to Eq. (\ref{P_GAUSSIAN}), the averaged outcome $\bar{u}$ of $N$
binary measurements has the same conditional expectation value as that of each
binary measurement [Eq. (\ref{UA})], while its conditional fluctuation is
$\sqrt{N}$ times smaller [cf. Eq. (\ref{SA})], consistent with the central
limit theorem. As a result, the distinguishability (denoted by $\mathcal{D})$
between $\mathcal{P}(\bar{u}|\pm\alpha)$ or equivalently the strength of the
multi-outcome measurement is $\sqrt{N}$ times that of each binary
measurement:
\begin{equation}
\mathcal{D}(N)\equiv\sqrt{N}D. \label{DN}%
\end{equation}
Physically, under the QND condition, the eigenstates $|\pm\alpha\rangle$ of
the observable $\boldsymbol{\hat{\alpha}}\cdot\mathbf{\hat{I}}$ are
simultaneous eigenstates of $\{\hat{M}_{u}\}$ and $e^{-i\boldsymbol{\varphi
}\cdot\mathbf{\hat{I}}}$, so it remains invariant during the sequential
measurements. This allows $|\pm\alpha\rangle$ to be measured repeatedly to
improve the distinguishability between $\mathcal{P}(\bar{u}|\pm\alpha)$:\ the
signal-to-noise ratio provided by $\bar{u}$ -- the average of $N$ binary
outcomes -- is $\sqrt{N}$ times that of a single binary outcome $u$.

For sufficiently large $N$ and hence $D$, the two curves $\mathcal{P}(\bar
{u}|\pm\alpha)$ have negligible overlap, see Fig. \ref{G_DISTRIBUTION}(d) for
an example. Then an outcome $\bar{u}_{0}$ lies under either $\mathcal{P}%
(\bar{u}|\alpha)$ or $\mathcal{P}(\bar{u}|-\alpha)$, so a \textit{single}
outcome $\bar{u}_{0}$ is sufficient for a reliable discremination of
$|\pm\alpha\rangle$, i.e.,\ $\bar{u}_{0}$ under $\mathcal{P}(\bar{u}|a)$
($a=\pm\alpha$) indicates the initial state to be $|a\rangle$.
Correspondingly, the multi-outcome measurement becomes
projective:\ $\mathcal{\hat{M}}_{\bar{u}_{0}}\propto|a\rangle\langle a|$ for
$\bar{u}_{0}$ under $\mathcal{P}(\bar{u}|a)$.

\begin{figure}[ptb]
\includegraphics[width=\columnwidth]{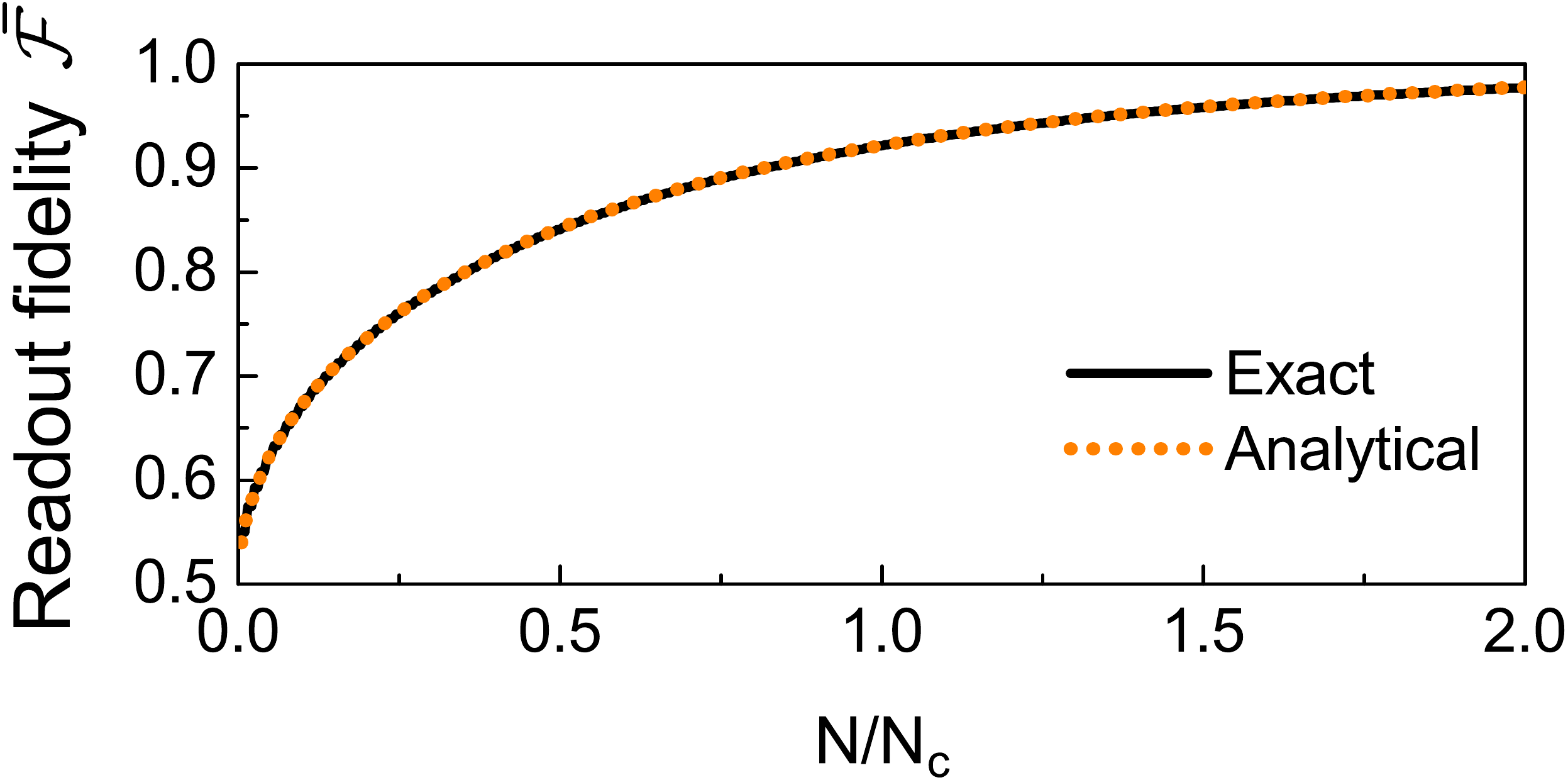}\caption{Average readout
fidelity $\mathcal{\bar{F}}$ of the multi-outcome measurement as a function of
the number $N$ (in units of $N_{\mathrm{c}}\equiv2/D^{2}$) of constituent
binary measurements. Here $\phi=\pi/2$ and $\alpha=0.1$, i.e., the strength of
each binary measurement is $D=0.1$.}%
\label{G_FIDELITY}%
\end{figure}

For finite $N$, the conditional distributions $\mathcal{P}(\bar{u}|\pm\alpha)$
always have a finite overlap. In this case, we can introduce a threshold
$\bar{u}_{\mathrm{th}}$ lying between $\langle u\rangle_{\alpha}$ and $\langle
u\rangle_{-\alpha}$ [see Fig. \ref{G_DISTRIBUTION}(c) for an example] and
identify the initial state as $|a\rangle$ ($a=\pm\alpha$) if $\bar{u}$ lies on
the side of $\langle u\rangle_{a}$. This identification is correct if the
outcome $\bar{u}$ lies outside the overlapping region, but it could be
incorrect if $\bar{u}$ lies inside the small overlapping region. To quantify
its accuracy, we define the readout\textit{ fidelity} $\mathcal{F}_{a}$ of the
state $|a\rangle$ ($a=\pm\alpha$) as the probability to identify the initial
state to be $|a\rangle$ when the true initial state is $|a\rangle$, e.g.,
$\mathcal{F}_{\alpha}=\sum_{\bar{u}>\bar{u}_{\mathrm{th}}}\mathcal{P}(\bar
{u}|\alpha)$ and $\mathcal{F}_{-\alpha}\equiv\sum_{\bar{u}<\bar{u}%
_{\mathrm{th}}}\mathcal{P}(\bar{u}|-\alpha)$ for $\langle u\rangle_{\alpha
}>\langle u\rangle_{-\alpha}$. Ideal projective measurement corresponds to
$\mathcal{F}_{\pm\alpha}=1$. Following Refs.
\cite{RobledoNature2011,DreauPRL2013}, we maximize the average readout
fidelity $\mathcal{\bar{F}}\equiv(\mathcal{F}_{\alpha}+\mathcal{F}_{-\alpha
})/2$ by setting $\bar{u}_{\mathrm{th}}$ at the overlapping point{, }see the
vertical dotted lines in Fig. \ref{G_DISTRIBUTION}(c). For $N\gg1$,
$\mathcal{P}(\bar{u}|\pm\alpha)$ are approximately Gaussian {[}Eq.
(\ref{P_GAUSSIAN}){]}, then
\[
\bar{u}_{\mathrm{th}}\approx\frac{\langle u\rangle_{\alpha}/\sigma_{\alpha
}+\langle u\rangle_{-\alpha}/\sigma_{-\alpha}}{1/\sigma_{\alpha}%
+1/\sigma_{-\alpha}},
\]
so the readout fidelity
\begin{equation}
\mathcal{\bar{F}}=\mathcal{F}_{\pm\alpha}\approx\frac{1}{2}+\frac{1}%
{2}\mathrm{erf}(\frac{\mathcal{D}}{\sqrt{2}}) \label{FBAR}%
\end{equation}
is a \textit{universal} function of the distinguishability $\mathcal{D}$ [Eq.
(\ref{DN})] between $\mathcal{P}(\bar{u}|\pm\alpha)$ or equivalently
measurement strength. As shown in Fig. \ref{G_FIDELITY}, Eq. (\ref{FBAR})
agrees well with the exact results. When $\mathcal{D}\gtrsim1$, $\mathcal{\bar
{F}}$ approaches 100\%, so the multi-outcome measurement approaches an ideal
projective measurement. For example, $N=N_{\mathrm{c}}\equiv2/D^{2}$ leads to
$\mathcal{D}=\sqrt{2}$ and hence $\mathcal{\bar{F}}\approx92\%$, while
$N=2N_{\mathrm{c}}$ leads to $\mathcal{D}=2$ and hence $\mathcal{\bar{F}%
}\approx98\%$.

For clarity, in the following we define $\mathcal{\bar{F}}_{\mathrm{th}}%
\equiv92\%$ or equivalently $\mathcal{D}_{\mathrm{th}}\equiv\sqrt{2}$ as the
\textit{threshold} of a high-fidelity projective measurement.

\subsection{Stability analysis}

\label{SEC_STABILITY}

When the QND condition Eq. (\ref{QND}) is violated slightly, the rotation
$e^{-i\boldsymbol{\varphi}\cdot\mathbf{\hat{I}}}$ after each binary
measurement will rotate the eigenstates $|\pm\alpha\rangle$ of the observable
$\hat{\boldsymbol{\alpha}}\cdot\hat{\mathbf{I}}$ slightly away from the
measurement axis $\hat{\boldsymbol{\alpha}}$, then the initial state
$|\pm\alpha\rangle$ is destroyed after certain number (denoted by
$N_{\mathrm{L}}$ -- the \textquotedblleft lifetime\textquotedblright\ of
$|\pm\alpha\rangle$) of binary measurements. Then the initial state
$|\pm\alpha\rangle$ can be measured repeated by at most $N_{\mathrm{L}}$
binary measurements, so the strength of the resulting multi-outcome
measurement can reach at most $\mathcal{D}(N_{\mathrm{L}})=\sqrt
{N_{\mathrm{L}}}D$. Then constructing a projective measurement above the
threshold fidelity $\mathcal{\bar{F}}_{\mathrm{th}}\equiv92\%$ requires
$\mathcal{D}(N_{\mathrm{L}})\geq\mathcal{D}_{\mathrm{th}}\equiv\sqrt{2}$ or
equivalently long lifetime
\begin{equation}
N_{\mathrm{L}}\gtrsim N_{\mathrm{c}}\equiv\frac{2}{D^{2}}. \label{HFD}%
\end{equation}

For a quantitative discussion, we calculate the \textit{unconditional} (i.e.,
with the measurement outcomes discarded) survival probability of $|\pm
\alpha\rangle$\ after a sequence of $N$ binary measurements, with the $i$th
measurement followed by the rotation $e^{-i\delta\boldsymbol{\varphi}_{i}%
\cdot\mathbf{\hat{I}}}$. A binary measurement with outcome discarded changes a
general nuclear spin state $\hat{\rho}=1/2+\hat{\mathbf{I}}\cdot\mathbf{n}$
with polarization $\mathbf{n}$ into $\sum_{u}\hat{M}_{u}\hat{\rho}\hat{M}%
_{u}^{\dagger}=1/2+\hat{\mathbf{I}}\cdot\boldsymbol{\mathbb{M}}\mathbf{n}$,
where $\boldsymbol{\mathbb{M}}\equiv\lbrack\boldsymbol{\mathbb{R}%
}(\boldsymbol{\alpha})+\boldsymbol{\mathbb{R}}(-\boldsymbol{\alpha})]/2$
describes the measurement-induced dephasing and $\boldsymbol{\mathbb{R}%
}(\boldsymbol{\theta})$ is the SO(3) rotation matrix as defined after Eq.
(\ref{ALPHA}). The effect of $\boldsymbol{\mathbb{M}}$ is to reduce the
polarization components perpendicular to $\hat{\boldsymbol{\alpha}}$ by a
factor $\cos\alpha$. For clarity we take the initial state as $|\alpha\rangle
$. At the end of the evolution, the unconditional nuclear spin state
$\hat{\rho}(N)=1/2+\hat{\mathbf{I}}\cdot\hat{\boldsymbol{\alpha}}(N)$ is
characterized by the polarization
\[
\hat{\boldsymbol{\alpha}}(N)=\boldsymbol{\mathbb{R}}(\delta\boldsymbol{\varphi
}_{N})\boldsymbol{\mathbb{M}}\cdots\boldsymbol{\mathbb{R}}(\delta
\boldsymbol{\varphi}_{1})\boldsymbol{\mathbb{M}}\hat{\boldsymbol{\alpha}}%
\]
and the survival probability of $|\alpha\rangle$ is
\[
\langle\alpha|\hat{\rho}(N)|\alpha\rangle=\frac{1+\hat{\boldsymbol{\alpha}%
}\cdot\hat{\boldsymbol{\alpha}}(N)}{2}\equiv\frac{1+S(N)}{2}.
\]
For $|-\alpha\rangle$ as the initial state, we obtain the same survival
probability. We define the lifetime $N_{\mathrm{L}}$ of $|\pm\alpha\rangle$ as
the characteristic $N$ for $S(N)$ to decay to $1/e$. In general, numerical
calculations are necessary to determine $S(N)$ and hence $N_{\mathrm{L}}$.

The QND-breaking effect, i.e., the limitation to the lifetime $N_{\mathrm{L}}%
$, originates from the perpendicular component of $\{\delta\boldsymbol{\varphi
}_{i}\}$ with respect to the measurement axis $\hat{\boldsymbol{\alpha}}$. The
\textit{worst case} arises if $\{\delta\boldsymbol{\varphi}_{i}\}$ are all
along the same direction (so that different rotations $e^{-i\delta
\boldsymbol{\varphi}_{i}\cdot\mathbf{\hat{I}}}$ add up constructively) and
this direction is perpendicular to $\hat{\boldsymbol{\alpha}}$ (so that each
rotation rotates $|\pm\alpha\rangle$ away from the measurement axis
$\hat{\boldsymbol{\alpha}}$ most efficiently). Interestingly, in this worst
case, we can obtain analytical results. For clarity we define $\hat
{\boldsymbol{\alpha}}$ as the $+X$ axis and $\delta\boldsymbol{\varphi}_{i}$
as the $+Z$ axis, then$\ \delta\boldsymbol{\varphi}_{i}=\delta\varphi
_{i}\mathbf{e}_{Z}$. We begin with two special cases:
\[
S(N)=%
\begin{cases}
\cos(\sum\limits_{i=1}^{N}(-1)^{i}\delta\varphi_{i}) & \cos\alpha=-1,\\
\prod\limits_{i=1}^{N}\cos(\delta\varphi_{i}) & \cos\alpha=0.
\end{cases}
\]
The case $\cos\alpha=-1$ corresponds to $\boldsymbol{\mathbb{M}}%
=\boldsymbol{\mathbb{R}}(\pi\mathbf{e}_{X})$, i.e., each binary measurement
causes $\pi$-rotation of the nuclear spin around the $X$ axis. Then, two
binary measurements can reverse a rotation $\boldsymbol{\mathbb{R}}%
(\delta\varphi\mathbf{e}_{Z})$ around the $Z$ axis: $\boldsymbol{\mathbb{MR}%
}(\delta\varphi\mathbf{e}_{Z})\boldsymbol{\mathbb{M}}=\boldsymbol{\mathbb{R}%
}(-\delta\varphi\mathbf{e}_{Z})$. This measurement-induced spin echo can
suppress the decay of $S(N)$ when $\delta\varphi_{i}$ varies with $i$ slowly.
For $\cos\alpha=0$, the dephasing $\boldsymbol{\mathbb{M}}$ eliminates all the
$YZ$ components of the nuclear spin polarization, so the $i$th binary
measurement reduces the length of the polarization by a factor $\cos
(\delta\varphi_{i})$.

\begin{figure}[ptb]
\includegraphics[width=\columnwidth]{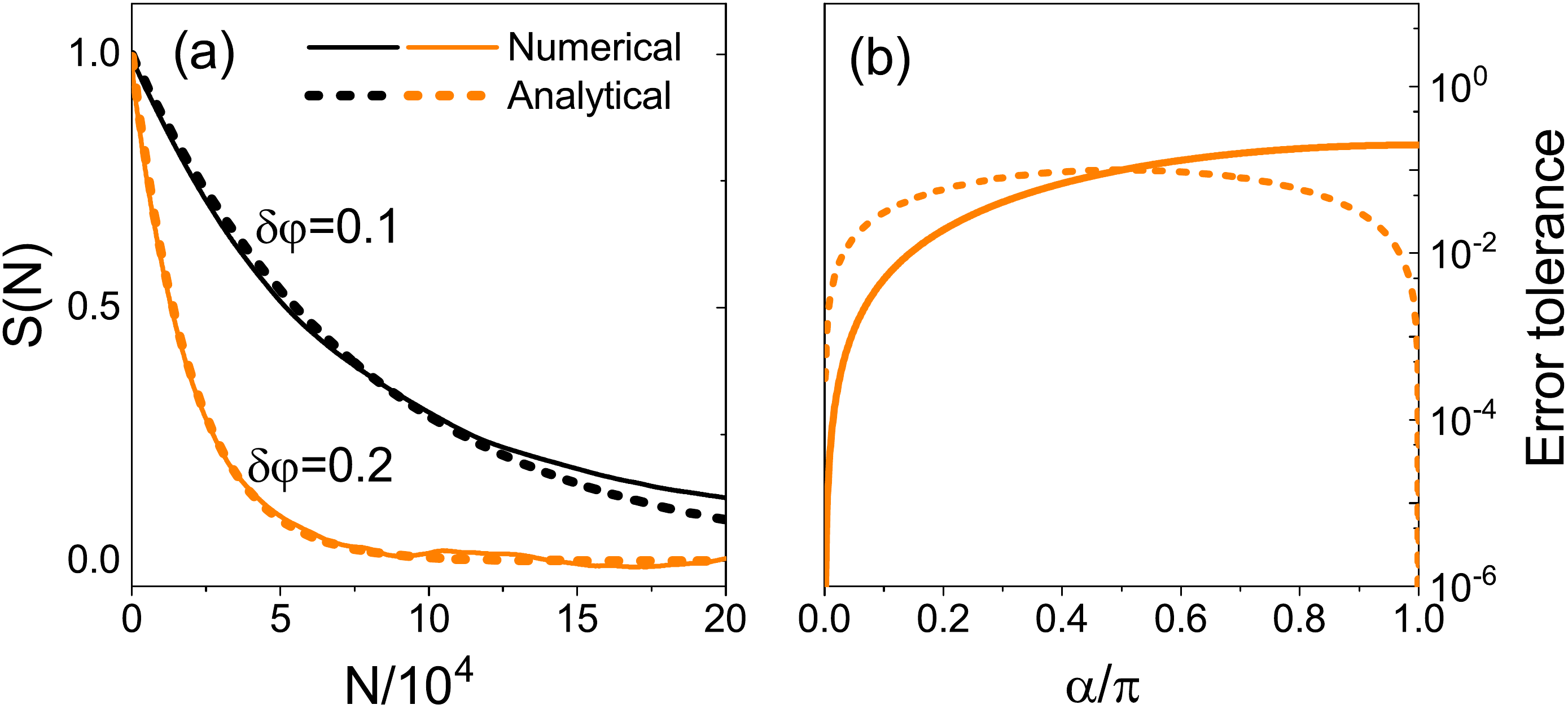}\caption{(a) Stability
$S(N)$ of $|\alpha\rangle$ against small systematic rotation error for
$\boldsymbol{\alpha}=(\pi-0.1)\mathbf{e}_{\pi/4}$ and $\phi=-\pi/2$. (b)
Tolerance against systematic error (solid line) and uncorrelated random error
(dashed line) for electron spin reaodut fidelity $p=0.1$.}%
\label{G_LIFETIME}%
\end{figure}

For small systematic rotation error $\delta\varphi_{i}=\delta\varphi\ll
\tan^{2}(\alpha/2)$, we obtain $S(N)\approx e^{-N(\delta\varphi)^{2}%
/[2\tan^{2}(\alpha/2)]}$, which agrees well with the numerical simulations
[see Fig. \ref{G_LIFETIME}(a)]. The lifetime
\begin{equation}
N_{\mathrm{L}}\approx\frac{2}{(\delta\varphi)^{2}}\tan^{2}\frac{\alpha}{2}
\label{NS_SYS}%
\end{equation}
depends strongly on $\alpha$ and hence the measurement strength of each binary
measurement. When the rotation errors $\{\delta\varphi_{i}\}$ are small
uncorrelated random numbers with standard deviation $\delta\varphi$, we obtain
$S(N)=e^{-N(\delta\varphi)^{2}/2}$, so the lifetime
\begin{equation}
N_{\mathrm{L}}=\frac{2}{(\delta\varphi)^{2}} \label{NS_RAN}%
\end{equation}
is independent of $\alpha$. The condition Eq. (\ref{HFD}) for constructing a
projective measurement above the threshold fidelity $\mathcal{\bar{F}%
}_{\mathrm{th}}\equiv92\%$ becomes $\delta\varphi\leq\Delta\varphi$, where%
\begin{equation}
\Delta\varphi\equiv\left\{
\begin{array}
[c]{ll}%
D\left\vert \tan(\alpha/2)\right\vert \ \  & (\text{systematic error)}\\
D & \text{(uncorrelated random error)}%
\end{array}
\right.  , \label{DPHI}%
\end{equation}
is the tolerence to rotation errors. As shown in Fig. \ref{G_LIFETIME}(b), for
small $\alpha$, the binary measurement strength $D\propto\alpha$, so the
tolerance against systematic (uncorrelated random) rotation error increases
quadratically (linearly) with $\alpha$. For $\alpha\rightarrow\pi$, the
measurement strength $D\rightarrow0$ linearly, so the tolerance against
uncorrelated random rotation error also approaches zero linearly. By contrast,
when $\alpha\rightarrow\pi$, the lifetime $N_{\mathrm{L}}$ due to systematic
rotation error diverges quadratically [see Eq. (\ref{NS_SYS})] due to the
measurement-induced spin echo, so the tolerance against systematic rotation
error approaches a constant $\Delta\varphi\approx2\bar{p}$ (for $\bar{p}\ll1$).

When the DD sequence is the repetition of an even-order concatenated DD, the
QND condition can be satisfied by tuning $t_{R}$ only, without flipping the
electron spin during the waiting time [see the discussions after Eq.
(\ref{QND})]. In this case, we have $\boldsymbol{\varphi}_{R}%
=(\boldsymbol{\omega}+\mathbf{A}/2)t_{R}$. Then the rotation error
$\delta\varphi\sim\delta\varphi_{R}=|\boldsymbol{\omega}+\mathbf{A}/2|\delta
t_{R}$ traces back to the error $\delta t_{R}$ in controlling the waiting
time. The tolerance against rotation error also translates to the tolerance
against the waiting time:%
\begin{equation}
\Delta t_{R}\sim\frac{\Delta\varphi}{|\boldsymbol{\omega}+\mathbf{A}/2|}.
\label{DTR}%
\end{equation}

\section{Example:\ nitrogen-vacancy center}%

\begin{table}[tbp] \centering
\begin{tabular}
[c]{llll}\hline\hline
Parameter sets & P1\ \ \ \ \ \ \  & P2\ \ \ \ \ \  & P3 \ \ \ \ \ \ \\
Number of CPMG period:$\ N_{\mathrm{DD}}$ & $6$ & $6$ & $8$\\
Magnetic field $B$\ \ \ \  & $691$ G & $305$ G & $305$ G\\
Larmor period $T_{R}\equiv2\pi/|\omega_{n}|$\ \ \ \ \ \ \  & $1351$ ns
\ \ \ \ \  & $3061$ ns\ \ \ \ \ \  & $3061$ ns\\
Larmor period $T\equiv2\pi/\left\vert \boldsymbol{\omega}\right\vert $ &
$1088$ ns & $1936$ ns & $1936$ ns\\\hline\hline
\end{tabular}
\caption{Three sets of $(N_{\mathrm{p}},B)$ parameters labelled by P1, P2, and P3. $T$ ($T_R$) is the period of the Larmor precession during the DD sequence (waiting time) in the weak hyperfine interaction approximation.}\label{PARAM}%
\end{table}%

We take an nitrogen-vacancy (NV) center electron spin-1 $\mathbf{\hat{S}%
}_{\mathrm{NV}}$ coupled to a $^{13}$C nuclear spin-1/2 $\hat{\mathbf{I}}$ via
the hyperfine interaction $\hat{S}_{\mathrm{NV}}^{z}\mathbf{A}\cdot
\hat{\mathbf{I}}$ as a paradigmatic physical system to illustrate our method.
Under an external magnetic field $B$ along the N-V symmetry axis (defined as
the $z$ axis), we can single out two NV electron spin states $|+_{z}%
\rangle\equiv|m_{S}=0\rangle$ and $|-_{z}\rangle\equiv|m_{S}=-1\rangle$ to
form the auxillary electron spin-1/2, e.g., $\hat{S}_{z}\equiv|+_{z}%
\rangle\langle+_{z}|-|-_{z}\rangle\langle-_{z}|$. Then, in the interaction
picture of the electron spin-1/2, we recover the total Hamiltonian in Eq.
(\ref{HAMIL}). The hyperfine-shifted nuclear Larmor frequency is
$\boldsymbol{\omega}\equiv\omega_{n}\mathbf{e}_{z}-\mathbf{A}/2$, where
$\omega_{n}\equiv\gamma_{n}B$ and $\gamma_{n}=-10.71$ MHz/T is the
gyromagnetic ratio of the $^{13}$C nucleus. Next, we follow the standard steps
to construct a projective QND measurement on the target $^{13}$C nucleus.

\subsection{Electron-mediated measurement on $^{13}$C nucleus}

We use the protocol in Fig. \ref{G_SETUP}(a) to construct a single binary
measurement on the $^{13}$C nucleus. We take the DD sequence as the
$N_{\mathrm{DD}}$-period CPMG sequence $(\tau/4$-$\pi$-$\tau/2$-$\pi$%
-$\tau/4)^{N_{\mathrm{DD}}}$ and set $\phi=\pi/2$ (i.e., we measure the
electron spin observable $\hat{S}_{y}$) to maximize the strength of each
binary measurement. The fidelity of the fluorescence-based readout of the NV
center electron spin is determined by the average photon number $n_{\pm}$ for
the electron spin state $|\pm_{z}\rangle$ or equivalently the average photon
number $\bar{n}=(n_{+}+n_{-})/2$ and the fluorescence contrast $C\equiv
(n_{+}-n_{-})/(n_{+}+n_{-})$. Room-temperature experiments have $n_{-}%
<n_{+}\ll1$, so nonzero (zero) photon detection corresponds to the outcome
$u=+$ ($u=-$). The readout fidelities for $|\pm_{z}\rangle$ are $p_{+}=n_{+}$
and $p_{-}=1-n_{-}$, then $\Delta p=2\bar{n}-1$ and $\bar{p}=2\bar{n}C\ll1$.
Then strength of each binary measurement follows from Eq. (\ref{DLOW}) as%
\begin{equation}
D\approx\sqrt{\bar{n}}C|\sin\alpha|\approx0.05\left\vert \sin\alpha\right\vert
, \label{DNV_RT}%
\end{equation}
where we have used typical values $n_{+}=0.1$ and $n_{-}=0.7n_{+}$ or
equivalently $\bar{n}=0.085$ and $C=0.18$ in the last step. At low
temperature, using resonant optical excitation
\cite{RobledoNature2011,PfaffNatPhys2013,CramerNC2016,KalbNC2016,ReisererPRX2016,YangNatPhoton2016}
gives $n_{+}\gg n_{-}$ and hence much higher readout fidelities on the NV
electron spin, then Eq. (\ref{DLOW}) gives%
\begin{equation}
D\sim\bar{p}|\sin\alpha|\approx0.9\left\vert \sin\alpha\right\vert ,
\label{DNV_LOWT}%
\end{equation}
where we have used typical values $p_{+}\approx0.89$ and $p_{-}\approx0.99$ in
the last step. Thus the measurement strength at low temperature is stronger
than that at room temperature by a factor $\sim18$. For specificity, here we
consider experiments at \textit{room temperature}.

\subsection{QND condition and controlling parameters}

\begin{figure*}[ptb]
\includegraphics[width=\textwidth]{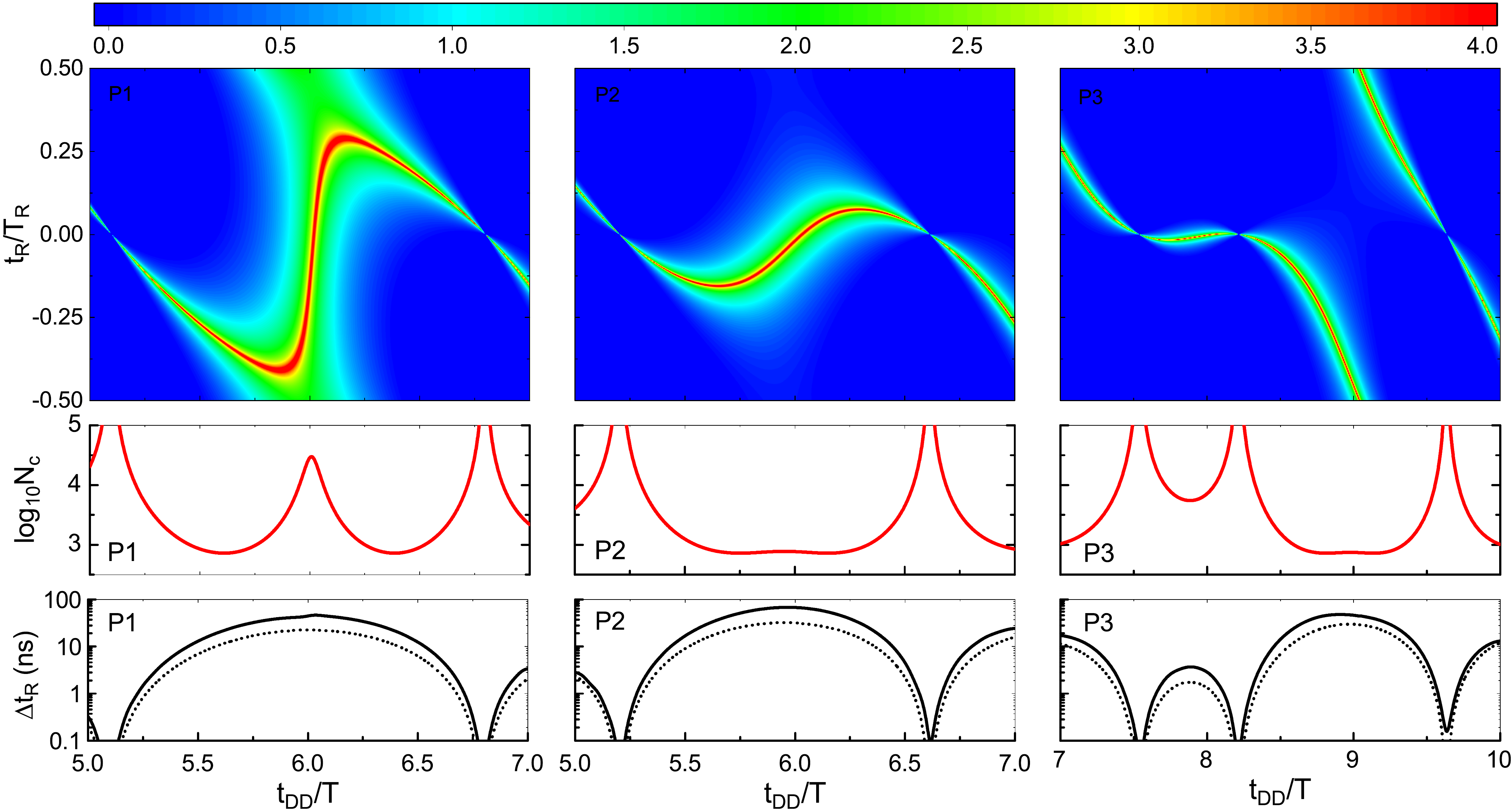}\caption{Upper pannels:\ contour
of $\log_{10}N_{\mathrm{L}}$ for P1-P3 in the $(t_{\mathrm{DD}}/T,t_{R}/T_{R}%
$) plane. Middle and lower panels: $\log_{10}N_{\mathrm{c}}$ and error
tolerance $\Delta t_{R}$ as function of $t_{\mathrm{DD}}/T$. The worst-case
estimation for $\Delta t_{R}$ is shown as the dashed\ lines.}%
\label{G_2DSCAN}%
\end{figure*}We use the protocol in Fig. \ref{G_SETUP}(c) to make each binary
measurement QND. According to the discussions after Eq. (\ref{QND}), after
measuring the electron spin, we immediately re-initialize the electron spin
into $|+_{z}\rangle=|m_{S}=0\rangle$ and then let the $^{13}$C nuclear spin
undergo free precession $e^{-i\boldsymbol{\varphi}_{R}\cdot\mathbf{\hat{I}}%
}=e^{-i\omega_{n}t_{R}\hat{I}_{z}}$ during the waiting time $t_{R}$. The total
rotation of the nuclear spin after the binary measurement is
$e^{-i\boldsymbol{\varphi}\cdot\mathbf{\hat{I}}}\equiv e^{-i\omega_{n}%
t_{R}\hat{I}_{z}}e^{-i\boldsymbol{\varphi}_{\mathrm{DD}}\cdot\mathbf{\hat{I}}%
}$. By varying $t_{R}$, we can tune $\boldsymbol{\varphi}$ towards the QND
condition Eq. (\ref{QND}) to provide a sufficiently long lifetime
$N_{\mathrm{L}}\geq N_{\mathrm{c}}$ [Eq. (\ref{HFD})], so that we can use
$N_{\mathrm{L}}$ repeated binary measurements to form a multi-outcome
projective measurement with readout fidelity above the threshold
$\mathcal{F}_{\mathrm{th}}\equiv92\%$.

The binary measurement is controlled by $\boldsymbol{\alpha}$, which depends
on the hyperfine interaction $\mathbf{A}$, the magnetic field $B$, the number
$N_{\mathrm{DD}}$ of CPMG periods, and the duration $\tau$ of each period. For
specificity, we set $\mathbf{A}=(0.316/\sqrt{2},0.316/\sqrt{2},0.330)$
$\mathrm{MHz}$ \cite{LiuPRL2017} with $1$ $\mathrm{MHz}\equiv2\pi\times10^{6}$
rad/(s T) and consider three sets of $(N_{\mathrm{DD}},B$), as labelled by P1,
P2, and P3 in Table \ref{PARAM}. For each set, we still have two controlling
parameters: the CPMG sequence period $\tau$ (or equivalently the CPMG sequence
duration $t_{\mathrm{DD}}\equiv N_{\mathrm{DD}}\tau$) and the waiting time
$t_{R}$. With $T\equiv2\pi/|\boldsymbol{\omega}|$ and $T_{R}\equiv2\pi
/|\omega_{n}|$ as the Larmor period of the target $^{13}$C nucleus during the
CPMG sequence and the waiting time, respectively, we can first tune $\tau$
close to resonance with $T$ (or equivalently $t_{\mathrm{DD}}$ close to
$N_{\mathrm{DD}}T$) to single out the target $^{13}$C from other environmental
nuclei and then tune $t_{R}$ (or \textit{fine} tune $\tau$) over one period
$[-T_{R}/2,T_{R}/2]$ to search for large $N_{\mathrm{L}}$. Next we perform a
numerical simulation for its error tolerance.

\subsection{Numerical simulation}

For each set of $(N_{\mathrm{DD}},B$), we scan $t_{R}$ over one period
$[-T_{R}/2,T_{R}/2]$ and scan $t_{\mathrm{DD}}\equiv N_{\mathrm{DD}}\tau$ (by
scanning $\tau$) in the vicinity of the resonance point $N_{\mathrm{DD}}T$.
The lifetime $N_{\mathrm{L}}$ for P1-P3 is shown in the upper panels of Fig.
\ref{G_2DSCAN}. The center of the red region correspond to diverging
$N_{\mathrm{L}}$ and hence the exact QND condition. Constructing a projective
measurement with readout fidelity above the threshold $\mathcal{\bar{F}%
}_{\mathrm{th}}\equiv92\%$ requires long lifetime $N_{\mathrm{L}}\geq
N_{\mathrm{c}}$ [Eq. (\ref{HFD})], where $N_{\mathrm{c}}$ as a function of
$t_{\mathrm{DD}}$ for P1-P3 is shown in the middle panel of Fig.
\ref{G_2DSCAN}. For each given $t_{\mathrm{DD}}$ in the $(t_{\mathrm{DD}%
},t_{R})$ plane, the width of the high-fidelity region (in which
$N_{\mathrm{L}}\geq N_{\mathrm{c}}$) along the $t_{R}$ axis, i.e., the
tolerance $\Delta t_{R}$ against systematic control error of $t_{R}$, is shown
in the lower panel of Fig. \ref{G_2DSCAN}. The worst-case estimation (dashed
lines)%
\[
\Delta t_{R}\sim\frac{T_{R}\ }{\pi}\sqrt{\bar{n}}C\sin^{2}\frac{\alpha}{2}%
\]
based on Eqs. (\ref{DPHI})-(\ref{DNV_RT}) shows qualitatively similar
dependences on the CPMG duration $t_{\mathrm{DD}}$ as the exact numerical
results, but it significantly underestimate $\Delta t_{R}$, in agreement with
our discussions in Sec. \ref{SEC_STABILITY}. The error tolerance $\Delta
t_{R}$ are on the nanoseconds time scale, within reach of typical experiments.
Therefore, constructing projective QND measurements from a sequence of binary
measurements is possible for P1-P3. Moreover, if we work at low temperatures
and use resonant optical excitation for high-fidelity readout of the NV
electron spin, then we can further enhance $D$ and hence the error tolerance
by a factor of $\sim18$.

\section{Conclusion}

We have developed a general theory for constructing projective quantum
nondemolition (QND) measurement on an arbitrary\ nuclear spin-1/2 by measuring
an axillary electron spin in generic electron-nuclear spin systems coupled via
hyperfine interaction. A distinguishing feature is that the QND observable is
\textit{not} conserved during the free Larmor precession of the nuclear spin
and can be tuned \textit{in situ}. The key idea consists of three steps.
First, suitable dynamical decoupling control on the electron is used to design
the electron-nuclear entanglement and hence select the nuclear spin observable
to be measured. Second, the nuclear spin evolution between neighboring
measurements is tuned to make the measurement QND. Finally, a sequence of such
measurements are cascaded into a projective QND measurement. We identify
tunable parameters to control the QND observable and further find optimal
parameters that stabilize the QND measurement against experimental control
errors. This work provides a paradigm for building up QND measurement on
non-conserved observables by a sequence of non-projective measurements in
hybrid qubit systems, which may be relevant to the state preparation, quantum
sensing, and quantum error correction via projective QND measurements. The
formalisms developed here can also be used to design other QND measurements
via more general quantum controls or study other measurement backaction effect
in NV center and other solid-state spin systems, such as semiconductor quantum
dots and phosphorus and bismuth donors in silicon.

\begin{acknowledgments}
P.W. is supported by the Talents Introduction Foundation of Beijing Normal University with Grant No.310432106.
W.Y. is supported by the NSAF grant in NSFC with grant No. U1930402.  P.W. and R.B.L. were
supported by the Hong Kong Research Grants Council - General Research
Fund Project 14300119.
We acknowledge the computational sup-port from the Beijing
Computational Science Research Center (CSRC).
\end{acknowledgments}

\textit{Note added----} Recently, Ref.\cite{MaarXiv2022} discuss how a sequence of  mutually commuting, normal POVM  (which is precisely the QND condition discussed in our paper) cascade into a projective measurement and provide a simple example based on a toy model. Here we focus on a realistic physical system -- electron-nuclear spin systems coupled through realistic hyperfine interaction. In this system, the electron-mediated measurement on the nuclear spin is  not QND, in construct to Ref. \cite{MaarXiv2022}. We show how to use suitable quantum control to engineer such non-QND measurements into QND ones, how to tune the QND observables and control the strength of the QND measurements, how a sequence of such QND measurements cascade into a projective measurement, and how the QND measurement is stabilized against realistic experimental control errors. We also give an explicit scheme for constructing a projective QND measurement on a 13C nuclear spin weakly coupled to a nitrogen-vacancy center electron spin in diamond.

\appendix

\section{Evolution during DD sequence}

We assume the DD sequence consists of $N$ $\pi$-pulses at $t_{1}\leq t_{2}%
\leq\cdots\leq t_{N}$. During the DD sequence, the total Hamiltonian in the
interaction picture of the auxiliary electron is
\begin{equation}
\hat{H}(t)=[\boldsymbol{\omega}+s(t)\hat{S}_{z}\mathbf{A}]\cdot\hat
{\mathbf{I}}, \label{HT}%
\end{equation}
where the DD modulation function $s(t)$ starts from $s(0)=+1$ and switches its
sign at $t_{1},t_{2},\cdots,t_{N}$ \cite{CywinskiPRB2008,YangRPP2017}. The
evolution operator during the DD sequence is $e^{-i\boldsymbol{\omega
}_{\mathrm{f}}\cdot\mathbf{\hat{I}(}t_{\mathrm{p}}-t_{N})}\cdots
e^{-i\boldsymbol{\omega}_{\pm}\cdot\mathbf{\hat{I}(}t_{3}-t_{2})}%
e^{-i\boldsymbol{\omega}_{\mp}\cdot\mathbf{\hat{I}(}t_{2}-t_{1})}%
e^{-i\boldsymbol{\omega}_{\pm}\cdot\hat{\mathbf{I}}t_{1}}$%
\begin{equation}
\hat{U}_{\mathrm{DD}}=e^{-i(\boldsymbol{\omega}+(-1)^{N}\hat{S}_{z}%
\mathbf{A})\cdot\hat{\mathbf{I}}}\cdots e^{-i(\boldsymbol{\omega}-\hat{S}%
_{z}\mathbf{A})\cdot\hat{\mathbf{I}}}e^{-i(\boldsymbol{\omega}+\hat{S}%
_{z}\mathbf{A})\cdot\hat{\mathbf{I}}}. \label{UDD_A}%
\end{equation}
We can expand $\hat{U}_{\mathrm{DD}}$ using the eigenstates $|\pm_{z}\rangle$
of $\hat{S}_{z}$ as
\[
\hat{U}_{\mathrm{DD}}=\hat{U}_{\mathrm{DD}}^{(+)}|+_{z}\rangle\langle
+_{z}|+\hat{U}_{\mathrm{DD}}^{(-)}|-_{z}\rangle\langle-_{z}|,
\]
where $\hat{U}_{\mathrm{DD}}^{(\pm)}=(\hat{U}_{\mathrm{DD}})_{\hat{S}%
_{z}\rightarrow\pm1/2}$ are nuclear spin evolution operators for the electron
spin initial state $|\pm_{z}\rangle$. Next we define $\boldsymbol{\alpha}$ and
$\boldsymbol{\varphi}_{\mathrm{DD}}$ via
\begin{align*}
e^{2i\boldsymbol{\alpha}\cdot\mathbf{\hat{I}}}  &  \equiv(\hat{U}%
_{\mathrm{DD}}^{(+)})^{\dagger}\hat{U}_{\mathrm{DD}}^{(-)},\\
e^{-i\boldsymbol{\varphi}_{\mathrm{DD}}\cdot\mathbf{\hat{I}}}  &  \equiv
\hat{U}_{\mathrm{DD}}^{(+)}e^{i\boldsymbol{\alpha}\cdot\mathbf{\hat{I}}}%
=\hat{U}_{\mathrm{DD}}^{(-)}e^{-i\boldsymbol{\alpha}\cdot\mathbf{\hat{I}}},
\end{align*}
then $\hat{U}_{\mathrm{DD}}^{(\pm)}=e^{-i\boldsymbol{\varphi}_{\mathrm{DD}%
}\cdot\mathbf{\hat{I}}}e^{\mp i\boldsymbol{\alpha}\cdot\mathbf{\hat{I}}}$ and
we obtain Eq. (\ref{UDD}) in the main text.

\section{Weak hyperfine interaction}

The evolution operator during the DD sequence [Eq. (\ref{UDD_A})] can be
written as
\[
\hat{U}_{\mathrm{DD}}=e^{-i\boldsymbol{\omega}\cdot\hat{\mathbf{I}%
}t_{\mathrm{DD}}}\mathcal{T}e^{-i\hat{S}_{z}\int_{0}^{t_{\mathrm{DD}}%
}s(t)\mathbf{A}\cdot\hat{\mathbf{I}}(t)dt^{\prime}},
\]
where $\mathcal{T}$ is the time-ordering superoperator and $\mathbf{\hat{I}%
}(t)\equiv e^{i\boldsymbol{\omega}\cdot\mathbf{\hat{I}}t}\mathbf{\hat{I}%
}e^{-i\boldsymbol{\omega}\cdot\mathbf{\hat{I}}t}$. When the perpendicular part
$\mathbf{A}_{\perp}$ of $\mathbf{A}$ with respect to $\boldsymbol{\omega}$ is
much smaller than the nuclear Zeeman splitting $|\boldsymbol{\omega}|$, we can
use the first-order Magus expansion to obtain
\[
\hat{U}_{\mathrm{DD}}\approx e^{-i\boldsymbol{\omega}\cdot\mathbf{\hat{I}%
}t_{\mathrm{DD}}}e^{-i\hat{S}_{z}\int_{0}^{t_{\mathrm{DD}}}s(t)\mathbf{A}%
\cdot\mathbf{\hat{I}}(t)dt}.
\]

For convenience we define $\boldsymbol{\omega}$ as the $z$ axis
($\boldsymbol{\omega}=\omega\mathbf{e}_{z}$) and decompose $\mathbf{A}$ into
$A_{z}\mathbf{e}_{z}$ and $\mathbf{A}_{\perp}=A_{\perp}(\cos\Phi\mathbf{e}%
_{x}+\sin\Phi\mathbf{e}_{y})=A_{\perp}\operatorname{Re}\mathbf{e}_{+}%
e^{-i\Phi}$ ($\mathbf{e}_{\pm}\equiv\mathbf{e}_{x}\pm i\mathbf{e}_{y}$),
then\ $\mathbf{A}\cdot\mathbf{\hat{I}}(t)=A_{z}\hat{I}_{z}+A_{\perp
}\mathbf{\hat{I}\cdot}\operatorname{Re}\mathbf{e}_{+}e^{-i\Phi}e^{i\omega t}$.
For a DD sequence of duration $t_{\mathrm{DD}}$, the modulation function must
satisfy $\int_{0}^{t_{\mathrm{DD}}}s(t)dt=0$, thus the term $A_{z}\hat{I}_{z}$
is averaged out, and
\[
\hat{U}_{\mathrm{DD}}\approx e^{-\omega t_{\mathrm{DD}}\hat{I}_{z}}%
e^{-i\hat{S}_{z}t_{\mathrm{DD}}A_{\perp}\mathbf{\hat{I}}\cdot\operatorname{Re}%
\mathbf{e}_{+}e^{-i\Phi}f_{\mathrm{DD}}}\newline\newline=e^{-it_{\mathrm{DD}%
}(\boldsymbol{\omega}\cdot\mathbf{\hat{I}})}e^{-i2\hat{S}_{z}%
\boldsymbol{\alpha}\cdot\mathbf{\hat{I}}},
\]
where $f_{\mathrm{DD}}$ is given by Eq. (\ref{FDD}) and $\boldsymbol{\alpha}$
is given by Eq. (\ref{ALPHA}).

\section{Stroboscopic QND condition for even-order concatenated DD sequences}

\label{APPEND_PHIR}

\begin{figure}[ptb]
\includegraphics[width=1\columnwidth]{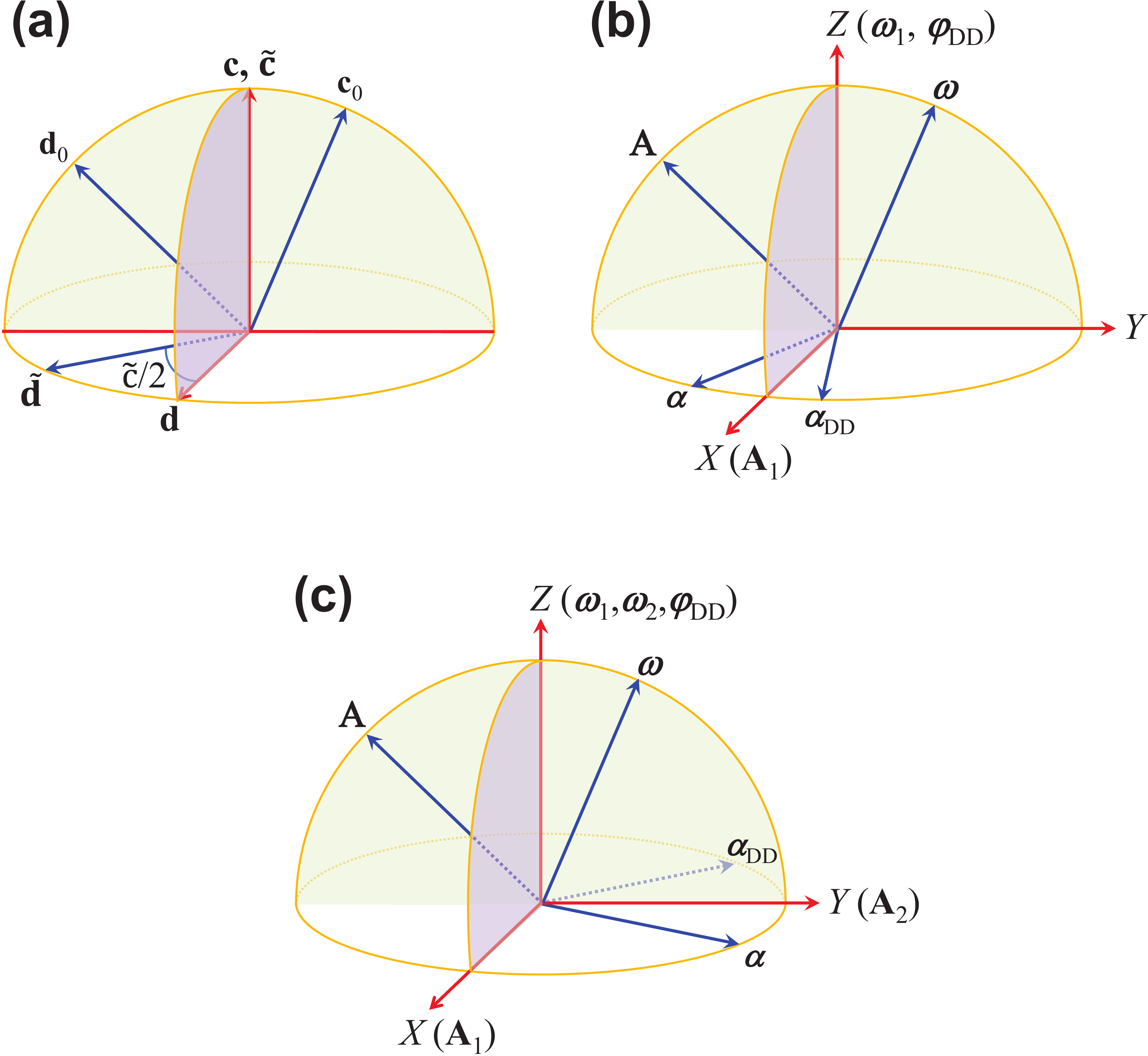}\caption{(a) Orientation
of relevant vectors for Eqs. (\ref{THEOREM1}) and (\ref{THEOREM2}). (b)
Orientations of relevant vectors for the periodic DD. (c) Orientations of
relevant vectors for the CPMG sequence. }%
\label{G_QND}%
\end{figure}

The QND condition Eq. (\ref{QND}) is equivalent to
\begin{equation}
\mathbb{R}(\boldsymbol{\varphi}_{R})\hat{\boldsymbol{\alpha}}_{\mathrm{DD}%
}=\hat{\boldsymbol{\alpha}}, \label{QND1}%
\end{equation}
where $\hat{\boldsymbol{\alpha}}_{\mathrm{DD}}\equiv\mathbb{R}%
(\boldsymbol{\varphi}_{\mathrm{DD}})\hat{\boldsymbol{\alpha}}$ is obtained
from $\hat{\boldsymbol{\alpha}}$ by a rotation around the axis
$\boldsymbol{\varphi}_{\mathrm{DD}}$ by an angle $|\boldsymbol{\varphi
}_{\mathrm{DD}}|$. This can be achieved by first tuning the direction of
$\boldsymbol{\varphi}_{R}$ into \textit{the bisection plane} of $\hat
{\boldsymbol{\alpha}}$ and $\hat{\boldsymbol{\alpha}}_{\mathrm{DD}}$ and then
tuning $|\boldsymbol{\varphi}_{R}|$ to satisfy Eq. (\ref{QND1}). As discussed
in the main text, we can achieve an \textit{arbitrary} evolution
$e^{-i\boldsymbol{\varphi}_{R}\cdot\mathbf{\hat{I}}}$ by tuning $t_{R}$ and
the timings of the electron spin flip during the waiting time $t_{R}$, so Eq.
(\ref{QND1}) can always be satisfied. Next, we further prove that when the DD
sequence is the repetition of an \textit{even-order} concatenated DD
\cite{KhodjastehPRL2005,YaoPRL2007,KhodjastehPRA2007,YangFP2011}, the QND
condition can be achieved by the evolution $e^{-i\boldsymbol{\varphi}_{R}%
\cdot\mathbf{\hat{I}}}=e^{-i(\boldsymbol{\omega}+\mathbf{A}/2)t_{R}%
\cdot\mathbf{\hat{I}}}$ at suitable $t_{R}$, i.e., for these DD sequences, the
QND condition can be satisfied by tuning the duration $t_{R}$ of the waiting
interval, without flipping the electron spin during the waiting interval.

We will use repeatedly two important properties (with $\hat{S}_{z}$ for the
electron spin-1/2 and $\mathbf{\hat{I}}$ for the target nuclear spin-1/2):
(i)\ Given arbitrary vectors $\mathbf{c}_{0}$ and $\mathbf{d}_{0}$, if we
define $\mathbf{c}$ and $\mathbf{d}$ via%
\begin{equation}
e^{-i(\mathbf{c}+\hat{S}_{z}\mathbf{d})\cdot\mathbf{\hat{I}}}=e^{-i\mathbf{(c}%
_{0}-\hat{S}_{z}\mathbf{d}_{0})\cdot\mathbf{\hat{I}}}e^{-i(\mathbf{c}_{0}%
+\hat{S}_{z}\mathbf{d}_{0})\cdot\mathbf{\hat{I}}}, \label{THEOREM1}%
\end{equation}
then $\mathbf{c}$ lies in the $\mathbf{c}_{0}$-$\mathbf{d}_{0}\ $plane
($\mathbf{c}\parallel\mathbf{c}_{0}$ if $\mathbf{c}_{0}\perp\mathbf{d}_{0}$),
while $\mathbf{d\parallel c}_{0}\times\mathbf{d}_{0}$, as shown in Fig.
\ref{G_QND}(a). (ii) Given two orthogonal vectors $\mathbf{c}\perp\mathbf{d}$,
if we define $\mathbf{\tilde{c}}$ and $\mathbf{\tilde{d}}$ via%
\begin{equation}
e^{-i\mathbf{\tilde{c}}\cdot\mathbf{\hat{I}}}e^{-i2\hat{S}_{z}\mathbf{\tilde
{d}}\cdot\mathbf{\hat{I}}}=e^{-i(\mathbf{c}+\hat{S}_{z}\mathbf{d}%
)\cdot\mathbf{\hat{I}}}, \label{THEOREM2}%
\end{equation}
then $\mathbf{\tilde{c}}\parallel\mathbf{c}$ and $\mathbf{\tilde{d}}%
\parallel\mathbb{R}(-\mathbf{\tilde{c}}/2)\mathbf{d}$, as shown in Fig.
\ref{G_QND}(a). The proof will be given at the end of this appendix.

To illustrate the concept of concatenation, we begin with the periodic DD
$(\tau/4$-$\pi$-$\tau/4$-$\pi$)$^{N}$ consisting of $N$ periods. The evolution
operator $\hat{U}_{1}$ of the electron-nuclear system in one period is the
concatenation of the free evolution $\hat{U}_{0}\equiv e^{-i(\tau
/4)(\boldsymbol{\omega}+\hat{S}_{z}\mathbf{A})\cdot\mathbf{\hat{I}}}$, i.e.,
\begin{equation}
\hat{U}_{1}\equiv(\hat{U}_{0})_{\hat{S}_{z}\rightarrow-\hat{S}_{z}}\hat{U}%
_{0}, \label{U1}%
\end{equation}
and the total evolution during this DD sequence is $\hat{U}_{\mathrm{DD}}%
=\hat{U}_{1}^{N}$. Namely, this periodic DD consists of $N$ repetitions of the
\textit{first-order} concatenated DD. Using Eq. (\ref{THEOREM1}), we have%
\begin{align*}
\hat{U}_{1}  &  =e^{-i(\tau/2)(\boldsymbol{\omega}_{1}+\hat{S}_{z}%
\mathbf{A}_{1}/2)\cdot\mathbf{\hat{I}}},\\
\hat{U}_{\mathrm{DD}}  &  =e^{-i(N\tau/2)(\boldsymbol{\omega}_{1}+\hat{S}%
_{z}\mathbf{A}_{1}/2)\cdot\mathbf{\hat{I}}},
\end{align*}
where $\boldsymbol{\omega}_{1}$ lies in the $\boldsymbol{\omega}$-$\mathbf{A}$
plane and $\mathbf{A}_{1}\parallel\boldsymbol{\omega}\times\mathbf{A}$. For
convenience, we define $\mathbf{A}_{1}$ as the $X$ axis, the
$\boldsymbol{\omega}$-$\mathbf{A}$ plane as the $YZ$ plane, and
$\boldsymbol{\omega}_{1}$ as the $Z$ axis, as shown in Fig. \ref{G_QND}(b).
Next we can use Eq. (\ref{THEOREM2}) to obtain $\hat{U}_{\mathrm{DD}%
}=e^{-i\boldsymbol{\varphi}_{\mathrm{DD}}\cdot\mathbf{\hat{I}}}e^{-i2\hat
{S}_{z}\boldsymbol{\alpha}\cdot\mathbf{\hat{I}}}$ with $\boldsymbol{\varphi
}_{\mathrm{DD}}$ along the $Z$ axis and $\boldsymbol{\hat{\alpha}}$ in the
$XY$ plane with azimuth $-|\boldsymbol{\varphi}_{\mathrm{DD}}|/2$, so
$\hat{\boldsymbol{\alpha}}_{\mathrm{DD}}$ also lies in the $XY$ plane with
azimuth $|\boldsymbol{\varphi}_{\mathrm{DD}}|/2$, i.e., $\boldsymbol{\hat
{\alpha}}$ and $\hat{\boldsymbol{\alpha}}_{\mathrm{DD}}$ lie symmetrically
about $\mathbf{A}_{1}$ in the $XY$ plane, as shown in Fig. \ref{G_QND}(b).
Therefore, for the periodic DD, the evolution $e^{-i\boldsymbol{\varphi}%
_{R}\cdot\mathbf{\hat{I}}}=e^{-i(\boldsymbol{\omega}+\mathbf{A}/2)t_{R}%
\cdot\mathbf{\hat{I}}}$ cannot satisfy the QND condition.

Next we consider the CPMG sequence $(\tau/4$-$\pi$-$\tau/2$-$\pi$-$\tau
/4$)$^{N}$ consisting of $N$ periods. The evolution operator of the
electron-nuclear system during one period is the concatenation of $\hat{U}%
_{1}$ [Eq. (\ref{U1})], i.e.,
\[
\hat{U}_{2}\equiv(\hat{U}_{1})_{\hat{S}_{z}\rightarrow-\hat{S}_{z}}\hat{U}%
_{1},
\]
and the total evolution during the CPMG sequence is $\hat{U}_{\mathrm{DD}%
}=\hat{U}_{2}^{N}$. Namely, this CPMG sequence consists of $N$ repetitions of
the \textit{second-order }concatenated DD. Using Eq. (\ref{THEOREM1}), we
obtain%
\[
\hat{U}_{2}=e^{-i\tau(\boldsymbol{\omega}_{2}+\hat{S}_{z}\mathbf{A}%
_{2}/2)\cdot\mathbf{\hat{I}}},
\]
with $\boldsymbol{\omega}_{2}$ along the $Z$ axis and $\mathbf{A}_{2}$ along
the $Y$ axis. Next we use Eq. (\ref{THEOREM2}) to obtain $\hat{U}%
_{\mathrm{DD}}=e^{-i\boldsymbol{\varphi}_{\mathrm{DD}}\cdot\mathbf{\hat{I}}%
}e^{-i2\hat{S}_{z}\boldsymbol{\alpha}\cdot\mathbf{\hat{I}}}$ with
$\boldsymbol{\varphi}_{\mathrm{DD}}$ along the $Z$ axis and $\boldsymbol{\hat
{\alpha}}$ in the $XY$ plane with azimuth $\pi/2-|\boldsymbol{\varphi
}_{\mathrm{DD}}|/2$, so $\hat{\boldsymbol{\alpha}}_{\mathrm{DD}}$ also lies in
the $XY$ plane with azimuth $\pi/2+|\boldsymbol{\varphi}_{\mathrm{DD}}|/2$,
i.e., $\boldsymbol{\hat{\alpha}}$ and $\hat{\boldsymbol{\alpha}}_{\mathrm{DD}%
}$ lie symmetrically about $\mathbf{A}_{2}$ in the $XY$ plane [see Fig.
\ref{G_QND}(c)]. Then the evolution $e^{-i\boldsymbol{\varphi}_{R}%
\cdot\mathbf{\hat{I}}}=e^{-i(\boldsymbol{\omega}+\mathbf{A}/2)\cdot
\mathbf{\hat{I}}t_{R}}$ can satisfy the stroboscopic QND condition at suitable
$t_{R}$, i.e., without flipping the electron spin during the waiting time.

The concatenation process can be carried out to higher order to obtain the
period
\[
\hat{U}_{l}\equiv(\hat{U}_{l-1})_{\hat{S}_{z}\rightarrow-\hat{S}_{z}}\hat
{U}_{l-1}%
\]
of the $l$th-order concatenated DD
\cite{KhodjastehPRL2005,YaoPRL2007,KhodjastehPRA2007,YangFP2011}. If the DD
sequence consists of $N$ repetitions of $\hat{U}_{l}$, then the total
evolution is $\hat{U}_{\mathrm{DD}}=\hat{U}_{l}^{N}$. Using Eq.
(\ref{THEOREM1}), we obtain
\[
\hat{U}_{l}=e^{-i2^{l-2}N\tau(\boldsymbol{\omega}_{l}+\hat{S}_{z}%
\mathbf{A}_{l})\cdot\mathbf{\hat{I}}},
\]
with $\boldsymbol{\omega}_{l}$ along the $Z$ axis and $\mathbf{A}_{l}$ in the
$XY$ plane with azimuth $(l-1)\pi/2$. Next we can use Eq. (\ref{THEOREM2}) to
obtain $\hat{U}_{\mathrm{DD}}=e^{-i\boldsymbol{\varphi}_{\mathrm{DD}}%
\cdot\mathbf{\hat{I}}}e^{-i2\hat{S}_{z}\boldsymbol{\alpha}\cdot\mathbf{\hat
{I}}}$, where $\boldsymbol{\varphi}_{\mathrm{DD}}$ is along the $Z$ axis and
$\boldsymbol{\hat{\alpha}}$ in the $XY$ plane with azimuth $(l-1)\pi
/2-|\boldsymbol{\varphi}_{\mathrm{DD}}|/2$, so $\hat{\boldsymbol{\alpha}%
}_{\mathrm{DD}}$ also lies in the $XY$ plane with azimuth $(l-1)\pi
/2+|\boldsymbol{\varphi}_{\mathrm{DD}}|/2$, i.e., $\boldsymbol{\hat{\alpha}}$
and $\hat{\boldsymbol{\alpha}}_{\mathrm{DD}}$ lie symmetrically about
$\mathbf{A}_{l}$ in the $XY$ plane. For \textit{even} $l$, $\mathbf{A}_{l}$ is
along the $\pm Y$ axis, so the evolution $e^{-i\boldsymbol{\varphi}_{R}%
\cdot\mathbf{\hat{I}}}=e^{-i(\boldsymbol{\omega}+\mathbf{A}/2)\cdot
\mathbf{\hat{I}}t_{R}}$ can satisfy the stroboscopic QND condition at suitable
$t_{R}$, i.e., without flipping the electron spin during the waiting time.

Finally we prove properties (i) and (ii). For property (i), we notice that Eq.
(\ref{THEOREM1}) is equivalent to%
\[
e^{-i(\mathbf{c}\pm\mathbf{d}/2)\cdot\mathbf{\hat{I}}}=e^{-i\mathbf{(c}_{0}%
\mp\mathbf{d}_{0}/2)\cdot\mathbf{\hat{I}}}e^{-i(\mathbf{c}_{0}\pm
\mathbf{d}_{0}/2)\cdot\mathbf{\hat{I}}},
\]
which further becomes
\[
e^{-i(\mathbf{C}\pm\mathbf{D})\cdot\boldsymbol{\sigma}}=e^{-i\mathbf{a}_{\mp
}\cdot\boldsymbol{\sigma}}e^{-i\mathbf{a}_{\pm}\cdot\boldsymbol{\sigma}}%
\]
by using $\mathbf{\hat{I}}=\boldsymbol{\sigma}/2$ ($\boldsymbol{\sigma}$ are
Pauli matrices) and defining $\mathbf{C}\equiv\mathbf{c}/2,$ $\mathbf{D}%
=\mathbf{d}/4$, and $\mathbf{a}_{\pm}\equiv(\mathbf{c}_{0}\pm\mathbf{d}%
_{0}/2)/2$. Using $e^{-i\theta(\mathbf{e}\cdot\boldsymbol{\sigma})}=\cos
\theta-i\boldsymbol{\sigma}\cdot\sin(\theta\mathbf{e})$ [$\mathbf{e}$ is a
unit vector and $\sin(\theta\mathbf{e})\equiv\mathbf{e}\sin\theta$], we obtain%
\begin{align*}
\cos\left\vert \mathbf{C}\pm\mathbf{D}\right\vert  &  =\cos a_{+}\cos
a_{-}-\sin\mathbf{a}_{+}\cdot\sin\mathbf{a}_{-},\\
\sin(\mathbf{C}\pm\mathbf{D})  &  =\cos a_{+}\sin\mathbf{a}_{-}+\sin
\mathbf{a}_{+}\cos a_{-}\pm\sin\mathbf{a}_{-}\mathbf{\times}\sin\mathbf{a}%
_{+},
\end{align*}
where $a_{\pm}=|\mathbf{a}_{\pm}|$. The first equation dictates $\left\vert
\mathbf{C}+\mathbf{D}\right\vert =\left\vert \mathbf{C}-\mathbf{D}\right\vert
$, then we can substitute $\sin(\mathbf{C}\pm\mathbf{D})=(\mathbf{C}%
\pm\mathbf{D})\operatorname{sinc}\sqrt{C^{2}+D^{2}}$ into the second equation
to obtain $\mathbf{C\propto}\sin\mathbf{a}_{-}\cos a_{+}+\sin\mathbf{a}%
_{+}\cos a_{-}$ and $\mathbf{D}\propto\sin\mathbf{a}_{-}\times\sin
\mathbf{a}_{+}$. In other words, $\mathbf{C}$ lies in the $\mathbf{a}_{+}%
$-$\mathbf{a}_{-}\ $plane ($\mathbf{C}\parallel\mathbf{a}_{+}+\mathbf{a}_{-}$
if $a_{+}=a_{-}$), while $\mathbf{D\parallel a}_{-}\times\mathbf{a}_{+}$. This
proves property (i).

Similarly, Eq. (\ref{THEOREM2}) is equivalent to%
\[
e^{-i\mathbf{\tilde{c}}\cdot\mathbf{\hat{I}}}e^{\mp i\mathbf{\tilde{d}}%
\cdot\mathbf{\hat{I}}}=e^{-i(\mathbf{c}\pm\mathbf{d}/2)\cdot\mathbf{\hat{I}}%
},
\]
which further becomes%
\[
e^{-i\mathbf{\tilde{C}}\cdot\boldsymbol{\sigma}}e^{\mp i\mathbf{\tilde{D}%
}\cdot\boldsymbol{\sigma}}=e^{-i(\mathbf{C}\pm\mathbf{D})\cdot
\boldsymbol{\sigma}}%
\]
with $\mathbf{C}=\mathbf{c}/2,$ $\mathbf{D}=\mathbf{d}/4$, $\mathbf{\tilde{C}%
}=\mathbf{\tilde{c}}/2$, and $\mathbf{\tilde{D}}=\mathbf{\tilde{d}}/2$. Using
$e^{-i\theta(\mathbf{e}\cdot\boldsymbol{\sigma})}=\cos\theta
-i\boldsymbol{\sigma}\cdot\sin(\theta\mathbf{e})$, we obtain%
\begin{align*}
\cos\sqrt{C^{2}+D^{2}}  &  =\cos\tilde{C}\cos\tilde{D}\mp\sin\mathbf{\tilde
{C}\cdot}\sin\mathbf{\tilde{D},}\\
(\mathbf{C}\pm\mathbf{D})\operatorname{sinc}\sqrt{C^{2}+D^{2}}  &
=\sin\mathbf{\tilde{C}}\cos\tilde{D}\pm\sin\mathbf{\tilde{D}}\cos\tilde{C}%
\pm\sin\mathbf{\tilde{C}\times}\sin\mathbf{\tilde{D}.}%
\end{align*}
The first equation dictates $\mathbf{\tilde{C}}\perp\mathbf{\tilde{D}}$. The
second equation gives $\mathbf{C}\propto\sin\mathbf{\tilde{C}}$ and
$\mathbf{D}\propto\sin\mathbf{\tilde{D}}\cos\tilde{C}+\sin\mathbf{\tilde
{C}\times}\sin\mathbf{\tilde{D}}=\mathbb{R}(\mathbf{\tilde{C}})\sin
\mathbf{\tilde{D}}$, so $\mathbf{\tilde{C}}\parallel\mathbf{C}$ and
$\mathbf{\tilde{D}}\parallel\mathbb{R}(-\mathbf{\tilde{C}})\mathbf{D}$. This
proves property (ii).

%



\begin{thebibliography}{52}%
\makeatletter
\providecommand \@ifxundefined [1]{%
 \@ifx{#1\undefined}
}%
\providecommand \@ifnum [1]{%
 \ifnum #1\expandafter \@firstoftwo
 \else \expandafter \@secondoftwo
 \fi
}%
\providecommand \@ifx [1]{%
 \ifx #1\expandafter \@firstoftwo
 \else \expandafter \@secondoftwo
 \fi
}%
\providecommand \natexlab [1]{#1}%
\providecommand \enquote  [1]{``#1''}%
\providecommand \bibnamefont  [1]{#1}%
\providecommand \bibfnamefont [1]{#1}%
\providecommand \citenamefont [1]{#1}%
\providecommand \href@noop [0]{\@secondoftwo}%
\providecommand \href [0]{\begingroup \@sanitize@url \@href}%
\providecommand \@href[1]{\@@startlink{#1}\@@href}%
\providecommand \@@href[1]{\endgroup#1\@@endlink}%
\providecommand \@sanitize@url [0]{\catcode `\\12\catcode `\$12\catcode
  `\&12\catcode `\#12\catcode `\^12\catcode `\_12\catcode `\%12\relax}%
\providecommand \@@startlink[1]{}%
\providecommand \@@endlink[0]{}%
\providecommand \url  [0]{\begingroup\@sanitize@url \@url }%
\providecommand \@url [1]{\endgroup\@href {#1}{\urlprefix }}%
\providecommand \urlprefix  [0]{URL }%
\providecommand \Eprint [0]{\href }%
\providecommand \doibase [0]{http://dx.doi.org/}%
\providecommand \selectlanguage [0]{\@gobble}%
\providecommand \bibinfo  [0]{\@secondoftwo}%
\providecommand \bibfield  [0]{\@secondoftwo}%
\providecommand \translation [1]{[#1]}%
\providecommand \BibitemOpen [0]{}%
\providecommand \bibitemStop [0]{}%
\providecommand \bibitemNoStop [0]{.\EOS\space}%
\providecommand \EOS [0]{\spacefactor3000\relax}%
\providecommand \BibitemShut  [1]{\csname bibitem#1\endcsname}%
\let\auto@bib@innerbib\@empty
\bibitem [{\citenamefont {Wiseman}\ and\ \citenamefont
  {Milburn}(2010)}]{WisemanBook2010}%
  \BibitemOpen
  \bibfield  {author} {\bibinfo {author} {\bibfnamefont {H.~M.}\ \bibnamefont
  {Wiseman}}\ and\ \bibinfo {author} {\bibfnamefont {G.~J.}\ \bibnamefont
  {Milburn}},\ }\href@noop {} {\emph {\bibinfo {title} {Quantum measurement and
  control}}}\ (\bibinfo  {publisher} {Cambridge University Press},\ \bibinfo
  {year} {2010})\BibitemShut {NoStop}%
\bibitem [{\citenamefont {Braginsky}\ and\ \citenamefont
  {Vorontsov}(1975)}]{BraginskiSPU1975}%
  \BibitemOpen
  \bibfield  {author} {\bibinfo {author} {\bibfnamefont {V.~B.}\ \bibnamefont
  {Braginsky}}\ and\ \bibinfo {author} {\bibfnamefont {Y.~I.}\ \bibnamefont
  {Vorontsov}},\ }\href@noop {} {\bibfield  {journal} {\bibinfo  {journal}
  {Soviet Physics Uspekhi}\ }\textbf {\bibinfo {volume} {17}},\ \bibinfo
  {pages} {644} (\bibinfo {year} {1975})}\BibitemShut {NoStop}%
\bibitem [{\citenamefont {Braginskii}\ \emph {et~al.}(1977)\citenamefont
  {Braginskii}, \citenamefont {Vorontsov},\ and\ \citenamefont
  {Khalili}}]{BraginskiiJETP1977}%
  \BibitemOpen
  \bibfield  {author} {\bibinfo {author} {\bibfnamefont {V.}~\bibnamefont
  {Braginskii}}, \bibinfo {author} {\bibfnamefont {Y.~I.}\ \bibnamefont
  {Vorontsov}}, \ and\ \bibinfo {author} {\bibfnamefont {F.~Y.}\ \bibnamefont
  {Khalili}},\ }\href@noop {} {\bibfield  {journal} {\bibinfo  {journal} {Sov.
  Phys. JETP}\ }\textbf {\bibinfo {volume} {46}},\ \bibinfo {pages} {705}
  (\bibinfo {year} {1977})}\BibitemShut {NoStop}%
\bibitem [{\citenamefont {Thorne}\ \emph {et~al.}(1978)\citenamefont {Thorne},
  \citenamefont {Drever}, \citenamefont {Caves}, \citenamefont {Zimmermann},\
  and\ \citenamefont {Sandberg}}]{ThornePRL1978}%
  \BibitemOpen
  \bibfield  {author} {\bibinfo {author} {\bibfnamefont {K.~S.}\ \bibnamefont
  {Thorne}}, \bibinfo {author} {\bibfnamefont {R.~W.~P.}\ \bibnamefont
  {Drever}}, \bibinfo {author} {\bibfnamefont {C.~M.}\ \bibnamefont {Caves}},
  \bibinfo {author} {\bibfnamefont {M.}~\bibnamefont {Zimmermann}}, \ and\
  \bibinfo {author} {\bibfnamefont {V.~D.}\ \bibnamefont {Sandberg}},\
  }\href@noop {} {\bibfield  {journal} {\bibinfo  {journal} {Phys. Rev. Lett.}\
  }\textbf {\bibinfo {volume} {40}},\ \bibinfo {pages} {667} (\bibinfo {year}
  {1978})}\BibitemShut {NoStop}%
\bibitem [{\citenamefont {Unruh}(1979)}]{UnruhPRD1979}%
  \BibitemOpen
  \bibfield  {author} {\bibinfo {author} {\bibfnamefont {W.~G.}\ \bibnamefont
  {Unruh}},\ }\href@noop {} {\bibfield  {journal} {\bibinfo  {journal} {Phys.
  Rev. D}\ }\textbf {\bibinfo {volume} {19}},\ \bibinfo {pages} {2888}
  (\bibinfo {year} {1979})}\BibitemShut {NoStop}%
\bibitem [{\citenamefont {Caves}\ \emph {et~al.}(1980)\citenamefont {Caves},
  \citenamefont {Thorne}, \citenamefont {Drever}, \citenamefont {Sandberg},\
  and\ \citenamefont {Zimmermann}}]{CavesRMP1980}%
  \BibitemOpen
  \bibfield  {author} {\bibinfo {author} {\bibfnamefont {C.~M.}\ \bibnamefont
  {Caves}}, \bibinfo {author} {\bibfnamefont {K.~S.}\ \bibnamefont {Thorne}},
  \bibinfo {author} {\bibfnamefont {R.~W.~P.}\ \bibnamefont {Drever}}, \bibinfo
  {author} {\bibfnamefont {V.~D.}\ \bibnamefont {Sandberg}}, \ and\ \bibinfo
  {author} {\bibfnamefont {M.}~\bibnamefont {Zimmermann}},\ }\href@noop {}
  {\bibfield  {journal} {\bibinfo  {journal} {Rev. Mod. Phys.}\ }\textbf
  {\bibinfo {volume} {52}},\ \bibinfo {pages} {341} (\bibinfo {year}
  {1980})}\BibitemShut {NoStop}%
\bibitem [{\citenamefont {Braginsky}\ \emph {et~al.}(1980)\citenamefont
  {Braginsky}, \citenamefont {Vorontsov},\ and\ \citenamefont
  {Thorne}}]{BraginskyScience1980}%
  \BibitemOpen
  \bibfield  {author} {\bibinfo {author} {\bibfnamefont {V.~B.}\ \bibnamefont
  {Braginsky}}, \bibinfo {author} {\bibfnamefont {Y.~I.}\ \bibnamefont
  {Vorontsov}}, \ and\ \bibinfo {author} {\bibfnamefont {K.~S.}\ \bibnamefont
  {Thorne}},\ }\href@noop {} {\bibfield  {journal} {\bibinfo  {journal}
  {Science}\ }\textbf {\bibinfo {volume} {209}},\ \bibinfo {pages} {547}
  (\bibinfo {year} {1980})}\BibitemShut {NoStop}%
\bibitem [{\citenamefont {Clerk}\ \emph {et~al.}(2010)\citenamefont {Clerk},
  \citenamefont {Devoret}, \citenamefont {Girvin}, \citenamefont {Marquardt},\
  and\ \citenamefont {Schoelkopf}}]{ClerkRMP2010}%
  \BibitemOpen
  \bibfield  {author} {\bibinfo {author} {\bibfnamefont {A.~A.}\ \bibnamefont
  {Clerk}}, \bibinfo {author} {\bibfnamefont {M.~H.}\ \bibnamefont {Devoret}},
  \bibinfo {author} {\bibfnamefont {S.~M.}\ \bibnamefont {Girvin}}, \bibinfo
  {author} {\bibfnamefont {F.}~\bibnamefont {Marquardt}}, \ and\ \bibinfo
  {author} {\bibfnamefont {R.~J.}\ \bibnamefont {Schoelkopf}},\ }\href@noop {}
  {\bibfield  {journal} {\bibinfo  {journal} {Rev. Mod. Phys.}\ }\textbf
  {\bibinfo {volume} {82}},\ \bibinfo {pages} {1155} (\bibinfo {year}
  {2010})}\BibitemShut {NoStop}%
\bibitem [{\citenamefont {Grangier}\ \emph {et~al.}(1998)\citenamefont
  {Grangier}, \citenamefont {Levenson},\ and\ \citenamefont
  {Poizat}}]{GrangierNature1998}%
  \BibitemOpen
  \bibfield  {author} {\bibinfo {author} {\bibfnamefont {P.}~\bibnamefont
  {Grangier}}, \bibinfo {author} {\bibfnamefont {J.~A.}\ \bibnamefont
  {Levenson}}, \ and\ \bibinfo {author} {\bibfnamefont {J.-P.}\ \bibnamefont
  {Poizat}},\ }\href@noop {} {\bibfield  {journal} {\bibinfo  {journal}
  {Nature}\ }\textbf {\bibinfo {volume} {396}},\ \bibinfo {pages} {537}
  (\bibinfo {year} {1998})}\BibitemShut {NoStop}%
\bibitem [{\citenamefont {Lupascu}\ \emph {et~al.}(2007)\citenamefont
  {Lupascu}, \citenamefont {Saito}, \citenamefont {Picot}, \citenamefont
  {de~Groot}, \citenamefont {Harmans},\ and\ \citenamefont
  {Mooij}}]{LupascuNatPhys2007}%
  \BibitemOpen
  \bibfield  {author} {\bibinfo {author} {\bibfnamefont {A.}~\bibnamefont
  {Lupascu}}, \bibinfo {author} {\bibfnamefont {S.}~\bibnamefont {Saito}},
  \bibinfo {author} {\bibfnamefont {T.}~\bibnamefont {Picot}}, \bibinfo
  {author} {\bibfnamefont {P.~C.}\ \bibnamefont {de~Groot}}, \bibinfo {author}
  {\bibfnamefont {C.~J. P.~M.}\ \bibnamefont {Harmans}}, \ and\ \bibinfo
  {author} {\bibfnamefont {J.~E.}\ \bibnamefont {Mooij}},\ }\href@noop {}
  {\bibfield  {journal} {\bibinfo  {journal} {Nat. Phys.}\ }\textbf {\bibinfo
  {volume} {3}},\ \bibinfo {pages} {119} (\bibinfo {year} {2007})}\BibitemShut
  {NoStop}%
\bibitem [{\citenamefont {Peil}\ and\ \citenamefont
  {Gabrielse}(1999)}]{PeilPRL1999}%
  \BibitemOpen
  \bibfield  {author} {\bibinfo {author} {\bibfnamefont {S.}~\bibnamefont
  {Peil}}\ and\ \bibinfo {author} {\bibfnamefont {G.}~\bibnamefont
  {Gabrielse}},\ }\href@noop {} {\bibfield  {journal} {\bibinfo  {journal}
  {Phys. Rev. Lett.}\ }\textbf {\bibinfo {volume} {83}},\ \bibinfo {pages}
  {1287} (\bibinfo {year} {1999})}\BibitemShut {NoStop}%
\bibitem [{\citenamefont {Laraoui}\ \emph {et~al.}(2013)\citenamefont
  {Laraoui}, \citenamefont {Dolde}, \citenamefont {Burk}, \citenamefont
  {Reinhard}, \citenamefont {Wrachtrup},\ and\ \citenamefont
  {Meriles}}]{LaraouiNatCommun2013}%
  \BibitemOpen
  \bibfield  {author} {\bibinfo {author} {\bibfnamefont {A.}~\bibnamefont
  {Laraoui}}, \bibinfo {author} {\bibfnamefont {F.}~\bibnamefont {Dolde}},
  \bibinfo {author} {\bibfnamefont {C.}~\bibnamefont {Burk}}, \bibinfo {author}
  {\bibfnamefont {F.}~\bibnamefont {Reinhard}}, \bibinfo {author}
  {\bibfnamefont {J.}~\bibnamefont {Wrachtrup}}, \ and\ \bibinfo {author}
  {\bibfnamefont {C.~A.}\ \bibnamefont {Meriles}},\ }\href {\doibase
  10.1038/ncomms2685} {\bibfield  {journal} {\bibinfo  {journal} {Nature
  Communications}\ }\textbf {\bibinfo {volume} {4}},\ \bibinfo {pages} {1651}
  (\bibinfo {year} {2013})}\BibitemShut {NoStop}%
\bibitem [{\citenamefont {Staudacher}\ \emph {et~al.}(2015)\citenamefont
  {Staudacher}, \citenamefont {Raatz}, \citenamefont {Pezzagna}, \citenamefont
  {Meijer}, \citenamefont {Reinhard}, \citenamefont {Meriles},\ and\
  \citenamefont {Wrachtrup}}]{StaudacherNC2015}%
  \BibitemOpen
  \bibfield  {author} {\bibinfo {author} {\bibfnamefont {T.}~\bibnamefont
  {Staudacher}}, \bibinfo {author} {\bibfnamefont {N.}~\bibnamefont {Raatz}},
  \bibinfo {author} {\bibfnamefont {S.}~\bibnamefont {Pezzagna}}, \bibinfo
  {author} {\bibfnamefont {J.}~\bibnamefont {Meijer}}, \bibinfo {author}
  {\bibfnamefont {F.}~\bibnamefont {Reinhard}}, \bibinfo {author}
  {\bibfnamefont {C.~A.}\ \bibnamefont {Meriles}}, \ and\ \bibinfo {author}
  {\bibfnamefont {J.}~\bibnamefont {Wrachtrup}},\ }\href@noop {} {\bibfield
  {journal} {\bibinfo  {journal} {Nat. Commun.}\ }\textbf {\bibinfo {volume}
  {6}},\  (\bibinfo {year} {2015})}\BibitemShut {NoStop}%
\bibitem [{\citenamefont {Mamin}\ \emph {et~al.}(2013)\citenamefont {Mamin},
  \citenamefont {Kim}, \citenamefont {Sherwood}, \citenamefont {Rettner},
  \citenamefont {Ohno}, \citenamefont {Awschalom},\ and\ \citenamefont
  {Rugar}}]{MaminScience2013}%
  \BibitemOpen
  \bibfield  {author} {\bibinfo {author} {\bibfnamefont {H.~J.}\ \bibnamefont
  {Mamin}}, \bibinfo {author} {\bibfnamefont {M.}~\bibnamefont {Kim}}, \bibinfo
  {author} {\bibfnamefont {M.~H.}\ \bibnamefont {Sherwood}}, \bibinfo {author}
  {\bibfnamefont {C.~T.}\ \bibnamefont {Rettner}}, \bibinfo {author}
  {\bibfnamefont {K.}~\bibnamefont {Ohno}}, \bibinfo {author} {\bibfnamefont
  {D.~D.}\ \bibnamefont {Awschalom}}, \ and\ \bibinfo {author} {\bibfnamefont
  {D.}~\bibnamefont {Rugar}},\ }\href@noop {} {\bibfield  {journal} {\bibinfo
  {journal} {Science}\ }\textbf {\bibinfo {volume} {339}},\ \bibinfo {pages}
  {557} (\bibinfo {year} {2013})}\BibitemShut {NoStop}%
\bibitem [{\citenamefont {Wang}\ \emph {et~al.}(2017)\citenamefont {Wang},
  \citenamefont {Casanova},\ and\ \citenamefont {Plenio}}]{WangNC2017}%
  \BibitemOpen
  \bibfield  {author} {\bibinfo {author} {\bibfnamefont {Z.-Y.}\ \bibnamefont
  {Wang}}, \bibinfo {author} {\bibfnamefont {J.}~\bibnamefont {Casanova}}, \
  and\ \bibinfo {author} {\bibfnamefont {M.~B.}\ \bibnamefont {Plenio}},\
  }\href@noop {} {\bibfield  {journal} {\bibinfo  {journal} {Nat. Commun.}\
  }\textbf {\bibinfo {volume} {8}},\ \bibinfo {pages} {14660} (\bibinfo {year}
  {2017})}\BibitemShut {NoStop}%
\bibitem [{\citenamefont {Zaiser}\ \emph {et~al.}(2016)\citenamefont {Zaiser},
  \citenamefont {Rendler}, \citenamefont {Jakobi}, \citenamefont {Wolf},
  \citenamefont {Lee}, \citenamefont {Wagner}, \citenamefont {Bergholm},
  \citenamefont {Schulte-Herbrüggen}, \citenamefont {Neumann},\ and\
  \citenamefont {Wrachtrup}}]{ZaiserNC2016}%
  \BibitemOpen
  \bibfield  {author} {\bibinfo {author} {\bibfnamefont {S.}~\bibnamefont
  {Zaiser}}, \bibinfo {author} {\bibfnamefont {T.}~\bibnamefont {Rendler}},
  \bibinfo {author} {\bibfnamefont {I.}~\bibnamefont {Jakobi}}, \bibinfo
  {author} {\bibfnamefont {T.}~\bibnamefont {Wolf}}, \bibinfo {author}
  {\bibfnamefont {S.-Y.}\ \bibnamefont {Lee}}, \bibinfo {author} {\bibfnamefont
  {S.}~\bibnamefont {Wagner}}, \bibinfo {author} {\bibfnamefont
  {V.}~\bibnamefont {Bergholm}}, \bibinfo {author} {\bibfnamefont
  {T.}~\bibnamefont {Schulte-Herbrüggen}}, \bibinfo {author} {\bibfnamefont
  {P.}~\bibnamefont {Neumann}}, \ and\ \bibinfo {author} {\bibfnamefont
  {J.}~\bibnamefont {Wrachtrup}},\ }\href {\doibase 10.1038/ncomms12279}
  {\bibfield  {journal} {\bibinfo  {journal} {Nature Communications}\ }\textbf
  {\bibinfo {volume} {7}},\ \bibinfo {pages} {12279} (\bibinfo {year}
  {2016})}\BibitemShut {NoStop}%
\bibitem [{\citenamefont {Shi}\ \emph {et~al.}(2014)\citenamefont {Shi},
  \citenamefont {Kong}, \citenamefont {Wang}, \citenamefont {Kong},
  \citenamefont {Zhao}, \citenamefont {Liu},\ and\ \citenamefont
  {Du}}]{ShiNatPhys2014}%
  \BibitemOpen
  \bibfield  {author} {\bibinfo {author} {\bibfnamefont {F.}~\bibnamefont
  {Shi}}, \bibinfo {author} {\bibfnamefont {X.}~\bibnamefont {Kong}}, \bibinfo
  {author} {\bibfnamefont {P.}~\bibnamefont {Wang}}, \bibinfo {author}
  {\bibfnamefont {F.}~\bibnamefont {Kong}}, \bibinfo {author} {\bibfnamefont
  {N.}~\bibnamefont {Zhao}}, \bibinfo {author} {\bibfnamefont {R.-B.}\
  \bibnamefont {Liu}}, \ and\ \bibinfo {author} {\bibfnamefont
  {J.}~\bibnamefont {Du}},\ }\href@noop {} {\bibfield  {journal} {\bibinfo
  {journal} {Nat. Phys.}\ }\textbf {\bibinfo {volume} {10}},\ \bibinfo {pages}
  {21} (\bibinfo {year} {2014})}\BibitemShut {NoStop}%
\bibitem [{\citenamefont {Shi}\ \emph {et~al.}(2015)\citenamefont {Shi},
  \citenamefont {Zhang}, \citenamefont {Wang}, \citenamefont {Sun},
  \citenamefont {Wang}, \citenamefont {Rong}, \citenamefont {Chen},
  \citenamefont {Ju}, \citenamefont {Friedemann}, \citenamefont {Wang},\ and\
  \citenamefont {Du}}]{ShiScience2015}%
  \BibitemOpen
  \bibfield  {author} {\bibinfo {author} {\bibfnamefont {F.}~\bibnamefont
  {Shi}}, \bibinfo {author} {\bibfnamefont {Q.}~\bibnamefont {Zhang}}, \bibinfo
  {author} {\bibfnamefont {P.}~\bibnamefont {Wang}}, \bibinfo {author}
  {\bibfnamefont {H.}~\bibnamefont {Sun}}, \bibinfo {author} {\bibfnamefont
  {J.}~\bibnamefont {Wang}}, \bibinfo {author} {\bibfnamefont {X.}~\bibnamefont
  {Rong}}, \bibinfo {author} {\bibfnamefont {M.}~\bibnamefont {Chen}}, \bibinfo
  {author} {\bibfnamefont {C.}~\bibnamefont {Ju}}, \bibinfo {author}
  {\bibfnamefont {R.}~\bibnamefont {Friedemann}}, \bibinfo {author}
  {\bibfnamefont {J.}~\bibnamefont {Wang}}, \ and\ \bibinfo {author}
  {\bibfnamefont {J.}~\bibnamefont {Du}},\ }\href@noop {} {\bibfield  {journal}
  {\bibinfo  {journal} {Science}\ }\textbf {\bibinfo {volume} {347}},\ \bibinfo
  {pages} {1135} (\bibinfo {year} {2015})}\BibitemShut {NoStop}%
\bibitem [{\citenamefont {Du}\ \emph {et~al.}(2009)\citenamefont {Du},
  \citenamefont {Rong}, \citenamefont {Zhao}, \citenamefont {Wang},
  \citenamefont {Yang},\ and\ \citenamefont {Liu}}]{DuNature2009}%
  \BibitemOpen
  \bibfield  {author} {\bibinfo {author} {\bibfnamefont {J.}~\bibnamefont
  {Du}}, \bibinfo {author} {\bibfnamefont {X.}~\bibnamefont {Rong}}, \bibinfo
  {author} {\bibfnamefont {N.}~\bibnamefont {Zhao}}, \bibinfo {author}
  {\bibfnamefont {Y.}~\bibnamefont {Wang}}, \bibinfo {author} {\bibfnamefont
  {J.}~\bibnamefont {Yang}}, \ and\ \bibinfo {author} {\bibfnamefont {R.~B.}\
  \bibnamefont {Liu}},\ }\href@noop {} {\bibfield  {journal} {\bibinfo
  {journal} {Nature}\ }\textbf {\bibinfo {volume} {461}},\ \bibinfo {pages}
  {1265} (\bibinfo {year} {2009})}\BibitemShut {NoStop}%
\bibitem [{\citenamefont {Pfender}\ \emph {et~al.}(2017)\citenamefont
  {Pfender}, \citenamefont {Aslam}, \citenamefont {Sumiya}, \citenamefont
  {Onoda}, \citenamefont {Neumann}, \citenamefont {Isoya}, \citenamefont
  {Meriles},\ and\ \citenamefont {Wrachtrup}}]{PfenderNC2017}%
  \BibitemOpen
  \bibfield  {author} {\bibinfo {author} {\bibfnamefont {M.}~\bibnamefont
  {Pfender}}, \bibinfo {author} {\bibfnamefont {N.}~\bibnamefont {Aslam}},
  \bibinfo {author} {\bibfnamefont {H.}~\bibnamefont {Sumiya}}, \bibinfo
  {author} {\bibfnamefont {S.}~\bibnamefont {Onoda}}, \bibinfo {author}
  {\bibfnamefont {P.}~\bibnamefont {Neumann}}, \bibinfo {author} {\bibfnamefont
  {J.}~\bibnamefont {Isoya}}, \bibinfo {author} {\bibfnamefont {C.~A.}\
  \bibnamefont {Meriles}}, \ and\ \bibinfo {author} {\bibfnamefont
  {J.}~\bibnamefont {Wrachtrup}},\ }\href@noop {} {\bibfield  {journal}
  {\bibinfo  {journal} {Nat. Commun.}\ }\textbf {\bibinfo {volume} {8}},\
  \bibinfo {pages} {834} (\bibinfo {year} {2017})}\BibitemShut {NoStop}%
\bibitem [{\citenamefont {Rosskopf}\ \emph {et~al.}(2017)\citenamefont
  {Rosskopf}, \citenamefont {Zopes}, \citenamefont {Boss},\ and\ \citenamefont
  {Degen}}]{RosskopfQF2017}%
  \BibitemOpen
  \bibfield  {author} {\bibinfo {author} {\bibfnamefont {T.}~\bibnamefont
  {Rosskopf}}, \bibinfo {author} {\bibfnamefont {J.}~\bibnamefont {Zopes}},
  \bibinfo {author} {\bibfnamefont {J.~M.}\ \bibnamefont {Boss}}, \ and\
  \bibinfo {author} {\bibfnamefont {C.~L.}\ \bibnamefont {Degen}},\ }\href@noop
  {} {\bibfield  {journal} {\bibinfo  {journal} {npj Quantum Inf.}\ }\textbf
  {\bibinfo {volume} {3}},\ \bibinfo {pages} {33} (\bibinfo {year}
  {2017})}\BibitemShut {NoStop}%
\bibitem [{\citenamefont {Schmitt}\ \emph {et~al.}(2017)\citenamefont
  {Schmitt}, \citenamefont {Gefen}, \citenamefont {St{\"u}rner}, \citenamefont
  {Unden}, \citenamefont {Wolff}, \citenamefont {M{\"u}ller}, \citenamefont
  {Scheuer}, \citenamefont {Naydenov}, \citenamefont {Markham}, \citenamefont
  {Pezzagna}, \citenamefont {Meijer}, \citenamefont {Schwarz}, \citenamefont
  {Plenio}, \citenamefont {Retzker}, \citenamefont {McGuinness},\ and\
  \citenamefont {Jelezko}}]{SchmittScience2017}%
  \BibitemOpen
  \bibfield  {author} {\bibinfo {author} {\bibfnamefont {S.}~\bibnamefont
  {Schmitt}}, \bibinfo {author} {\bibfnamefont {T.}~\bibnamefont {Gefen}},
  \bibinfo {author} {\bibfnamefont {F.~M.}\ \bibnamefont {St{\"u}rner}},
  \bibinfo {author} {\bibfnamefont {T.}~\bibnamefont {Unden}}, \bibinfo
  {author} {\bibfnamefont {G.}~\bibnamefont {Wolff}}, \bibinfo {author}
  {\bibfnamefont {C.}~\bibnamefont {M{\"u}ller}}, \bibinfo {author}
  {\bibfnamefont {J.}~\bibnamefont {Scheuer}}, \bibinfo {author} {\bibfnamefont
  {B.}~\bibnamefont {Naydenov}}, \bibinfo {author} {\bibfnamefont
  {M.}~\bibnamefont {Markham}}, \bibinfo {author} {\bibfnamefont
  {S.}~\bibnamefont {Pezzagna}}, \bibinfo {author} {\bibfnamefont
  {J.}~\bibnamefont {Meijer}}, \bibinfo {author} {\bibfnamefont
  {I.}~\bibnamefont {Schwarz}}, \bibinfo {author} {\bibfnamefont
  {M.}~\bibnamefont {Plenio}}, \bibinfo {author} {\bibfnamefont
  {A.}~\bibnamefont {Retzker}}, \bibinfo {author} {\bibfnamefont {L.~P.}\
  \bibnamefont {McGuinness}}, \ and\ \bibinfo {author} {\bibfnamefont
  {F.}~\bibnamefont {Jelezko}},\ }\href@noop {} {\bibfield  {journal} {\bibinfo
   {journal} {Science}\ }\textbf {\bibinfo {volume} {356}},\ \bibinfo {pages}
  {832} (\bibinfo {year} {2017})}\BibitemShut {NoStop}%
\bibitem [{\citenamefont {Glenn}\ \emph {et~al.}(2018)\citenamefont {Glenn},
  \citenamefont {Bucher}, \citenamefont {Lee}, \citenamefont {Lukin},
  \citenamefont {Park},\ and\ \citenamefont {Walsworth}}]{GlennNature2018}%
  \BibitemOpen
  \bibfield  {author} {\bibinfo {author} {\bibfnamefont {D.~R.}\ \bibnamefont
  {Glenn}}, \bibinfo {author} {\bibfnamefont {D.~B.}\ \bibnamefont {Bucher}},
  \bibinfo {author} {\bibfnamefont {J.}~\bibnamefont {Lee}}, \bibinfo {author}
  {\bibfnamefont {M.~D.}\ \bibnamefont {Lukin}}, \bibinfo {author}
  {\bibfnamefont {H.}~\bibnamefont {Park}}, \ and\ \bibinfo {author}
  {\bibfnamefont {R.~L.}\ \bibnamefont {Walsworth}},\ }\href {\doibase
  10.1038/nature25781} {\bibfield  {journal} {\bibinfo  {journal} {Nature}\
  }\textbf {\bibinfo {volume} {555}},\ \bibinfo {pages} {351} (\bibinfo {year}
  {2018})}\BibitemShut {NoStop}%
\bibitem [{\citenamefont {Muhonen}\ \emph {et~al.}(2014)\citenamefont
  {Muhonen}, \citenamefont {Dehollain}, \citenamefont {Laucht}, \citenamefont
  {Hudson}, \citenamefont {Kalra}, \citenamefont {Sekiguchi}, \citenamefont
  {Itoh}, \citenamefont {Jamieson}, \citenamefont {McCallum}, \citenamefont
  {Dzurak},\ and\ \citenamefont {Morello}}]{MuhonenNatNano2014}%
  \BibitemOpen
  \bibfield  {author} {\bibinfo {author} {\bibfnamefont {J.~T.}\ \bibnamefont
  {Muhonen}}, \bibinfo {author} {\bibfnamefont {J.~P.}\ \bibnamefont
  {Dehollain}}, \bibinfo {author} {\bibfnamefont {A.}~\bibnamefont {Laucht}},
  \bibinfo {author} {\bibfnamefont {F.~E.}\ \bibnamefont {Hudson}}, \bibinfo
  {author} {\bibfnamefont {R.}~\bibnamefont {Kalra}}, \bibinfo {author}
  {\bibfnamefont {T.}~\bibnamefont {Sekiguchi}}, \bibinfo {author}
  {\bibfnamefont {K.~M.}\ \bibnamefont {Itoh}}, \bibinfo {author}
  {\bibfnamefont {D.~N.}\ \bibnamefont {Jamieson}}, \bibinfo {author}
  {\bibfnamefont {J.~C.}\ \bibnamefont {McCallum}}, \bibinfo {author}
  {\bibfnamefont {A.~S.}\ \bibnamefont {Dzurak}}, \ and\ \bibinfo {author}
  {\bibfnamefont {A.}~\bibnamefont {Morello}},\ }\href@noop {} {\bibfield
  {journal} {\bibinfo  {journal} {Nat Nano}\ }\textbf {\bibinfo {volume} {9}},\
  \bibinfo {pages} {986} (\bibinfo {year} {2014})}\BibitemShut {NoStop}%
\bibitem [{\citenamefont {Saeedi}\ \emph {et~al.}(2013)\citenamefont {Saeedi},
  \citenamefont {Simmons}, \citenamefont {Salvail}, \citenamefont {Dluhy},
  \citenamefont {Riemann}, \citenamefont {Abrosimov}, \citenamefont {Becker},
  \citenamefont {Pohl}, \citenamefont {Morton},\ and\ \citenamefont
  {Thewalt}}]{SaeediScience2013}%
  \BibitemOpen
  \bibfield  {author} {\bibinfo {author} {\bibfnamefont {K.}~\bibnamefont
  {Saeedi}}, \bibinfo {author} {\bibfnamefont {S.}~\bibnamefont {Simmons}},
  \bibinfo {author} {\bibfnamefont {J.~Z.}\ \bibnamefont {Salvail}}, \bibinfo
  {author} {\bibfnamefont {P.}~\bibnamefont {Dluhy}}, \bibinfo {author}
  {\bibfnamefont {H.}~\bibnamefont {Riemann}}, \bibinfo {author} {\bibfnamefont
  {N.~V.}\ \bibnamefont {Abrosimov}}, \bibinfo {author} {\bibfnamefont
  {P.}~\bibnamefont {Becker}}, \bibinfo {author} {\bibfnamefont {H.-J.}\
  \bibnamefont {Pohl}}, \bibinfo {author} {\bibfnamefont {J.~J.}\ \bibnamefont
  {Morton}}, \ and\ \bibinfo {author} {\bibfnamefont {M.~L.}\ \bibnamefont
  {Thewalt}},\ }\href@noop {} {\bibfield  {journal} {\bibinfo  {journal}
  {Science}\ }\textbf {\bibinfo {volume} {342}},\ \bibinfo {pages} {830}
  (\bibinfo {year} {2013})}\BibitemShut {NoStop}%
\bibitem [{\citenamefont {Tyryshkin}\ \emph {et~al.}(2012)\citenamefont
  {Tyryshkin}, \citenamefont {Tojo}, \citenamefont {Morton}, \citenamefont
  {Riemann}, \citenamefont {Abrosimov}, \citenamefont {Becker}, \citenamefont
  {Pohl}, \citenamefont {Schenkel}, \citenamefont {Thewalt}, \citenamefont
  {Itoh},\ and\ \citenamefont {Lyon}}]{TyryshkinNatMater2012}%
  \BibitemOpen
  \bibfield  {author} {\bibinfo {author} {\bibfnamefont {A.~M.}\ \bibnamefont
  {Tyryshkin}}, \bibinfo {author} {\bibfnamefont {S.}~\bibnamefont {Tojo}},
  \bibinfo {author} {\bibfnamefont {J.~J.~L.}\ \bibnamefont {Morton}}, \bibinfo
  {author} {\bibfnamefont {H.}~\bibnamefont {Riemann}}, \bibinfo {author}
  {\bibfnamefont {N.~V.}\ \bibnamefont {Abrosimov}}, \bibinfo {author}
  {\bibfnamefont {P.}~\bibnamefont {Becker}}, \bibinfo {author} {\bibfnamefont
  {H.-J.}\ \bibnamefont {Pohl}}, \bibinfo {author} {\bibfnamefont
  {T.}~\bibnamefont {Schenkel}}, \bibinfo {author} {\bibfnamefont {M.~L.~W.}\
  \bibnamefont {Thewalt}}, \bibinfo {author} {\bibfnamefont {K.~M.}\
  \bibnamefont {Itoh}}, \ and\ \bibinfo {author} {\bibfnamefont {S.~A.}\
  \bibnamefont {Lyon}},\ }\href@noop {} {\bibfield  {journal} {\bibinfo
  {journal} {Nat. Mater.}\ }\textbf {\bibinfo {volume} {11}},\ \bibinfo {pages}
  {143} (\bibinfo {year} {2012})}\BibitemShut {NoStop}%
\bibitem [{\citenamefont {Press}\ \emph {et~al.}(2010)\citenamefont {Press},
  \citenamefont {De~Greve}, \citenamefont {McMahon}, \citenamefont {Ladd},
  \citenamefont {Friess}, \citenamefont {Schneider}, \citenamefont {Kamp},
  \citenamefont {Hofling}, \citenamefont {Forchel},\ and\ \citenamefont
  {Yamamoto}}]{PressNatPhys2010}%
  \BibitemOpen
  \bibfield  {author} {\bibinfo {author} {\bibfnamefont {D.}~\bibnamefont
  {Press}}, \bibinfo {author} {\bibfnamefont {K.}~\bibnamefont {De~Greve}},
  \bibinfo {author} {\bibfnamefont {P.~L.}\ \bibnamefont {McMahon}}, \bibinfo
  {author} {\bibfnamefont {T.~D.}\ \bibnamefont {Ladd}}, \bibinfo {author}
  {\bibfnamefont {B.}~\bibnamefont {Friess}}, \bibinfo {author} {\bibfnamefont
  {C.}~\bibnamefont {Schneider}}, \bibinfo {author} {\bibfnamefont
  {M.}~\bibnamefont {Kamp}}, \bibinfo {author} {\bibfnamefont {S.}~\bibnamefont
  {Hofling}}, \bibinfo {author} {\bibfnamefont {A.}~\bibnamefont {Forchel}}, \
  and\ \bibinfo {author} {\bibfnamefont {Y.}~\bibnamefont {Yamamoto}},\
  }\href@noop {} {\bibfield  {journal} {\bibinfo  {journal} {Nat. Photon.}\
  }\textbf {\bibinfo {volume} {4}},\ \bibinfo {pages} {367} (\bibinfo {year}
  {2010})}\BibitemShut {NoStop}%
\bibitem [{\citenamefont {Taminiau}\ \emph {et~al.}(2014)\citenamefont
  {Taminiau}, \citenamefont {Cramer}, \citenamefont {van~der Sar},
  \citenamefont {Dobrovitski},\ and\ \citenamefont
  {Hanson}}]{TaminiauNatNano2014}%
  \BibitemOpen
  \bibfield  {author} {\bibinfo {author} {\bibfnamefont {T.~H.}\ \bibnamefont
  {Taminiau}}, \bibinfo {author} {\bibfnamefont {J.}~\bibnamefont {Cramer}},
  \bibinfo {author} {\bibfnamefont {T.}~\bibnamefont {van~der Sar}}, \bibinfo
  {author} {\bibfnamefont {V.~V.}\ \bibnamefont {Dobrovitski}}, \ and\ \bibinfo
  {author} {\bibfnamefont {R.}~\bibnamefont {Hanson}},\ }\href@noop {}
  {\bibfield  {journal} {\bibinfo  {journal} {Nat Nano}\ }\textbf {\bibinfo
  {volume} {9}},\ \bibinfo {pages} {171} (\bibinfo {year} {2014})}\BibitemShut
  {NoStop}%
\bibitem [{\citenamefont {Yang}\ \emph {et~al.}(2016)\citenamefont {Yang},
  \citenamefont {Wang}, \citenamefont {Rao}, \citenamefont {Hien~Tran},
  \citenamefont {Momenzadeh}, \citenamefont {Markham}, \citenamefont
  {Twitchen}, \citenamefont {Wang}, \citenamefont {Yang}, \citenamefont
  {Stöhr}, \citenamefont {Neumann}, \citenamefont {Kosaka},\ and\
  \citenamefont {Wrachtrup}}]{YangNatPhoton2016}%
  \BibitemOpen
  \bibfield  {author} {\bibinfo {author} {\bibfnamefont {S.}~\bibnamefont
  {Yang}}, \bibinfo {author} {\bibfnamefont {Y.}~\bibnamefont {Wang}}, \bibinfo
  {author} {\bibfnamefont {D.~D.~B.}\ \bibnamefont {Rao}}, \bibinfo {author}
  {\bibfnamefont {T.}~\bibnamefont {Hien~Tran}}, \bibinfo {author}
  {\bibfnamefont {A.~S.}\ \bibnamefont {Momenzadeh}}, \bibinfo {author}
  {\bibfnamefont {M.}~\bibnamefont {Markham}}, \bibinfo {author} {\bibfnamefont
  {D.~J.}\ \bibnamefont {Twitchen}}, \bibinfo {author} {\bibfnamefont
  {P.}~\bibnamefont {Wang}}, \bibinfo {author} {\bibfnamefont {W.}~\bibnamefont
  {Yang}}, \bibinfo {author} {\bibfnamefont {R.}~\bibnamefont {Stöhr}},
  \bibinfo {author} {\bibfnamefont {P.}~\bibnamefont {Neumann}}, \bibinfo
  {author} {\bibfnamefont {H.}~\bibnamefont {Kosaka}}, \ and\ \bibinfo {author}
  {\bibfnamefont {J.}~\bibnamefont {Wrachtrup}},\ }\href {\doibase
  10.1038/nphoton.2016.103} {\bibfield  {journal} {\bibinfo  {journal} {Nature
  Photonics}\ }\textbf {\bibinfo {volume} {10}},\ \bibinfo {pages} {507}
  (\bibinfo {year} {2016})}\BibitemShut {NoStop}%
\bibitem [{\citenamefont {Dreau}\ \emph {et~al.}(2013)\citenamefont {Dreau},
  \citenamefont {Spinicelli}, \citenamefont {Maze}, \citenamefont {Roch},\ and\
  \citenamefont {Jacques}}]{DreauPRL2013}%
  \BibitemOpen
  \bibfield  {author} {\bibinfo {author} {\bibfnamefont {A.}~\bibnamefont
  {Dreau}}, \bibinfo {author} {\bibfnamefont {P.}~\bibnamefont {Spinicelli}},
  \bibinfo {author} {\bibfnamefont {J.~R.}\ \bibnamefont {Maze}}, \bibinfo
  {author} {\bibfnamefont {J.-F.}\ \bibnamefont {Roch}}, \ and\ \bibinfo
  {author} {\bibfnamefont {V.}~\bibnamefont {Jacques}},\ }\href@noop {}
  {\bibfield  {journal} {\bibinfo  {journal} {Phys. Rev. Lett.}\ }\textbf
  {\bibinfo {volume} {110}},\ \bibinfo {pages} {060502} (\bibinfo {year}
  {2013})}\BibitemShut {NoStop}%
\bibitem [{\citenamefont {Neumann}\ \emph {et~al.}(2010)\citenamefont
  {Neumann}, \citenamefont {Beck}, \citenamefont {Steiner}, \citenamefont
  {Rempp}, \citenamefont {Fedder}, \citenamefont {Hemmer}, \citenamefont
  {Wrachtrup},\ and\ \citenamefont {Jelezko}}]{NeumannScience2010}%
  \BibitemOpen
  \bibfield  {author} {\bibinfo {author} {\bibfnamefont {P.}~\bibnamefont
  {Neumann}}, \bibinfo {author} {\bibfnamefont {J.}~\bibnamefont {Beck}},
  \bibinfo {author} {\bibfnamefont {M.}~\bibnamefont {Steiner}}, \bibinfo
  {author} {\bibfnamefont {F.}~\bibnamefont {Rempp}}, \bibinfo {author}
  {\bibfnamefont {H.}~\bibnamefont {Fedder}}, \bibinfo {author} {\bibfnamefont
  {P.~R.}\ \bibnamefont {Hemmer}}, \bibinfo {author} {\bibfnamefont
  {J.}~\bibnamefont {Wrachtrup}}, \ and\ \bibinfo {author} {\bibfnamefont
  {F.}~\bibnamefont {Jelezko}},\ }\href@noop {} {\bibfield  {journal} {\bibinfo
   {journal} {Science}\ }\textbf {\bibinfo {volume} {329}},\ \bibinfo {pages}
  {542} (\bibinfo {year} {2010})}\BibitemShut {NoStop}%
\bibitem [{\citenamefont {Liu}\ \emph {et~al.}(2017)\citenamefont {Liu},
  \citenamefont {Xing}, \citenamefont {Ma}, \citenamefont {Wang}, \citenamefont
  {Li}, \citenamefont {Po}, \citenamefont {Zhang}, \citenamefont {Fan},
  \citenamefont {Liu},\ and\ \citenamefont {Pan}}]{LiuPRL2017}%
  \BibitemOpen
  \bibfield  {author} {\bibinfo {author} {\bibfnamefont {G.-Q.}\ \bibnamefont
  {Liu}}, \bibinfo {author} {\bibfnamefont {J.}~\bibnamefont {Xing}}, \bibinfo
  {author} {\bibfnamefont {W.-L.}\ \bibnamefont {Ma}}, \bibinfo {author}
  {\bibfnamefont {P.}~\bibnamefont {Wang}}, \bibinfo {author} {\bibfnamefont
  {C.-H.}\ \bibnamefont {Li}}, \bibinfo {author} {\bibfnamefont {H.~C.}\
  \bibnamefont {Po}}, \bibinfo {author} {\bibfnamefont {Y.-R.}\ \bibnamefont
  {Zhang}}, \bibinfo {author} {\bibfnamefont {H.}~\bibnamefont {Fan}}, \bibinfo
  {author} {\bibfnamefont {R.-B.}\ \bibnamefont {Liu}}, \ and\ \bibinfo
  {author} {\bibfnamefont {X.-Y.}\ \bibnamefont {Pan}},\ }\href@noop {}
  {\bibfield  {journal} {\bibinfo  {journal} {Phys. Rev. Lett.}\ }\textbf
  {\bibinfo {volume} {118}},\ \bibinfo {pages} {150504} (\bibinfo {year}
  {2017})}\BibitemShut {NoStop}%
\bibitem [{\citenamefont {Braginsky}\ \emph {et~al.}(1978)\citenamefont
  {Braginsky}, \citenamefont {Vorontsov},\ and\ \citenamefont
  {Khalili}}]{BraginskyJETP1978}%
  \BibitemOpen
  \bibfield  {author} {\bibinfo {author} {\bibfnamefont {V.~B.}\ \bibnamefont
  {Braginsky}}, \bibinfo {author} {\bibfnamefont {Y.~I.}\ \bibnamefont
  {Vorontsov}}, \ and\ \bibinfo {author} {\bibfnamefont {F.~Y.}\ \bibnamefont
  {Khalili}},\ }\href@noop {} {\bibfield  {journal} {\bibinfo  {journal} {JETP
  Lett.}\ }\textbf {\bibinfo {volume} {27}},\ \bibinfo {pages} {276} (\bibinfo
  {year} {1978})}\BibitemShut {NoStop}%
\bibitem [{\citenamefont {Jordan}\ and\ \citenamefont
  {B\"uttiker}(2005)}]{JordanPRB2005}%
  \BibitemOpen
  \bibfield  {author} {\bibinfo {author} {\bibfnamefont {A.~N.}\ \bibnamefont
  {Jordan}}\ and\ \bibinfo {author} {\bibfnamefont {M.}~\bibnamefont
  {B\"uttiker}},\ }\href@noop {} {\bibfield  {journal} {\bibinfo  {journal}
  {Phys. Rev. B}\ }\textbf {\bibinfo {volume} {71}},\ \bibinfo {pages} {125333}
  (\bibinfo {year} {2005})}\BibitemShut {NoStop}%
\bibitem [{\citenamefont {Ruskov}\ \emph {et~al.}(2005)\citenamefont {Ruskov},
  \citenamefont {Schwab},\ and\ \citenamefont {Korotkov}}]{RuskovPRB2005a}%
  \BibitemOpen
  \bibfield  {author} {\bibinfo {author} {\bibfnamefont {R.}~\bibnamefont
  {Ruskov}}, \bibinfo {author} {\bibfnamefont {K.}~\bibnamefont {Schwab}}, \
  and\ \bibinfo {author} {\bibfnamefont {A.~N.}\ \bibnamefont {Korotkov}},\
  }\href@noop {} {\bibfield  {journal} {\bibinfo  {journal} {Phys. Rev. B}\
  }\textbf {\bibinfo {volume} {71}},\ \bibinfo {pages} {235407} (\bibinfo
  {year} {2005})}\BibitemShut {NoStop}%
\bibitem [{\citenamefont {Jordan}\ \emph {et~al.}(2006)\citenamefont {Jordan},
  \citenamefont {Korotkov},\ and\ \citenamefont {B\"uttiker}}]{JordanPRL2006}%
  \BibitemOpen
  \bibfield  {author} {\bibinfo {author} {\bibfnamefont {A.~N.}\ \bibnamefont
  {Jordan}}, \bibinfo {author} {\bibfnamefont {A.~N.}\ \bibnamefont
  {Korotkov}}, \ and\ \bibinfo {author} {\bibfnamefont {M.}~\bibnamefont
  {B\"uttiker}},\ }\href@noop {} {\bibfield  {journal} {\bibinfo  {journal}
  {Phys. Rev. Lett.}\ }\textbf {\bibinfo {volume} {97}},\ \bibinfo {pages}
  {026805} (\bibinfo {year} {2006})}\BibitemShut {NoStop}%
\bibitem [{\citenamefont {Averin}\ \emph {et~al.}(2006)\citenamefont {Averin},
  \citenamefont {Rabenstein},\ and\ \citenamefont {Semenov}}]{AverinPRB2006}%
  \BibitemOpen
  \bibfield  {author} {\bibinfo {author} {\bibfnamefont {D.~V.}\ \bibnamefont
  {Averin}}, \bibinfo {author} {\bibfnamefont {K.}~\bibnamefont {Rabenstein}},
  \ and\ \bibinfo {author} {\bibfnamefont {V.~K.}\ \bibnamefont {Semenov}},\
  }\href@noop {} {\bibfield  {journal} {\bibinfo  {journal} {Phys. Rev. B}\
  }\textbf {\bibinfo {volume} {73}},\ \bibinfo {pages} {094504} (\bibinfo
  {year} {2006})}\BibitemShut {NoStop}%
\bibitem [{\citenamefont {Jordan}\ and\ \citenamefont
  {Korotkov}(2006)}]{JordanPRB2006}%
  \BibitemOpen
  \bibfield  {author} {\bibinfo {author} {\bibfnamefont {A.~N.}\ \bibnamefont
  {Jordan}}\ and\ \bibinfo {author} {\bibfnamefont {A.~N.}\ \bibnamefont
  {Korotkov}},\ }\href@noop {} {\bibfield  {journal} {\bibinfo  {journal}
  {Phys. Rev. B}\ }\textbf {\bibinfo {volume} {74}},\ \bibinfo {pages} {085307}
  (\bibinfo {year} {2006})}\BibitemShut {NoStop}%
\bibitem [{\citenamefont {Greiner}\ \emph {et~al.}(2017)\citenamefont
  {Greiner}, \citenamefont {Dasari},\ and\ \citenamefont
  {Wrachtrup}}]{GreinerSciRep2017}%
  \BibitemOpen
  \bibfield  {author} {\bibinfo {author} {\bibfnamefont {J.~N.}\ \bibnamefont
  {Greiner}}, \bibinfo {author} {\bibfnamefont {D.~B.~R.}\ \bibnamefont
  {Dasari}}, \ and\ \bibinfo {author} {\bibfnamefont {J.}~\bibnamefont
  {Wrachtrup}},\ }\href@noop {} {\bibfield  {journal} {\bibinfo  {journal}
  {Sci. Rep.}\ }\textbf {\bibinfo {volume} {7}},\ \bibinfo {pages} {529}
  (\bibinfo {year} {2017})}\BibitemShut {NoStop}%
\bibitem [{\citenamefont {Cywinski}\ \emph {et~al.}(2008)\citenamefont
  {Cywinski}, \citenamefont {Lutchyn}, \citenamefont {Nave},\ and\
  \citenamefont {Das~Sarma}}]{CywinskiPRB2008}%
  \BibitemOpen
  \bibfield  {author} {\bibinfo {author} {\bibfnamefont {L.}~\bibnamefont
  {Cywinski}}, \bibinfo {author} {\bibfnamefont {R.~M.}\ \bibnamefont
  {Lutchyn}}, \bibinfo {author} {\bibfnamefont {C.~P.}\ \bibnamefont {Nave}}, \
  and\ \bibinfo {author} {\bibfnamefont {S.}~\bibnamefont {Das~Sarma}},\
  }\href@noop {} {\bibfield  {journal} {\bibinfo  {journal} {Phys. Rev. B}\
  }\textbf {\bibinfo {volume} {77}},\ \bibinfo {pages} {174509} (\bibinfo
  {year} {2008})}\BibitemShut {NoStop}%
\bibitem [{\citenamefont {Yang}\ \emph {et~al.}(2017)\citenamefont {Yang},
  \citenamefont {Ma},\ and\ \citenamefont {Liu}}]{YangRPP2017}%
  \BibitemOpen
  \bibfield  {author} {\bibinfo {author} {\bibfnamefont {W.}~\bibnamefont
  {Yang}}, \bibinfo {author} {\bibfnamefont {W.-L.}\ \bibnamefont {Ma}}, \ and\
  \bibinfo {author} {\bibfnamefont {R.-B.}\ \bibnamefont {Liu}},\ }\href@noop
  {} {\bibfield  {journal} {\bibinfo  {journal} {Rep. Prog. Phys.}\ }\textbf
  {\bibinfo {volume} {80}},\ \bibinfo {pages} {016001} (\bibinfo {year}
  {2017})}\BibitemShut {NoStop}%
\bibitem [{\citenamefont {Liu}\ \emph {et~al.}(2010)\citenamefont {Liu},
  \citenamefont {Yao},\ and\ \citenamefont {Sham}}]{LiuAdvPhys2010}%
  \BibitemOpen
  \bibfield  {author} {\bibinfo {author} {\bibfnamefont {R.-B.}\ \bibnamefont
  {Liu}}, \bibinfo {author} {\bibfnamefont {W.}~\bibnamefont {Yao}}, \ and\
  \bibinfo {author} {\bibfnamefont {L.~J.}\ \bibnamefont {Sham}},\ }\href@noop
  {} {\bibfield  {journal} {\bibinfo  {journal} {Adv. Phys.}\ }\textbf
  {\bibinfo {volume} {59}},\ \bibinfo {pages} {703} (\bibinfo {year}
  {2010})}\BibitemShut {NoStop}%
\bibitem [{\citenamefont {Khodjasteh}\ and\ \citenamefont
  {Lidar}(2005)}]{KhodjastehPRL2005}%
  \BibitemOpen
  \bibfield  {author} {\bibinfo {author} {\bibfnamefont {K.}~\bibnamefont
  {Khodjasteh}}\ and\ \bibinfo {author} {\bibfnamefont {D.~A.}\ \bibnamefont
  {Lidar}},\ }\href@noop {} {\bibfield  {journal} {\bibinfo  {journal} {Phys.
  Rev. Lett.}\ }\textbf {\bibinfo {volume} {95}},\ \bibinfo {pages} {180501}
  (\bibinfo {year} {2005})}\BibitemShut {NoStop}%
\bibitem [{\citenamefont {Yao}\ \emph {et~al.}(2007)\citenamefont {Yao},
  \citenamefont {Liu},\ and\ \citenamefont {Sham}}]{YaoPRL2007}%
  \BibitemOpen
  \bibfield  {author} {\bibinfo {author} {\bibfnamefont {W.}~\bibnamefont
  {Yao}}, \bibinfo {author} {\bibfnamefont {R.-B.}\ \bibnamefont {Liu}}, \ and\
  \bibinfo {author} {\bibfnamefont {L.~J.}\ \bibnamefont {Sham}},\ }\href@noop
  {} {\bibfield  {journal} {\bibinfo  {journal} {Phys. Rev. Lett.}\ }\textbf
  {\bibinfo {volume} {98}},\ \bibinfo {pages} {077602} (\bibinfo {year}
  {2007})}\BibitemShut {NoStop}%
\bibitem [{\citenamefont {Khodjasteh}\ and\ \citenamefont
  {Lidar}(2007)}]{KhodjastehPRA2007}%
  \BibitemOpen
  \bibfield  {author} {\bibinfo {author} {\bibfnamefont {K.}~\bibnamefont
  {Khodjasteh}}\ and\ \bibinfo {author} {\bibfnamefont {D.~A.}\ \bibnamefont
  {Lidar}},\ }\href@noop {} {\bibfield  {journal} {\bibinfo  {journal} {Phys.
  Rev. A}\ }\textbf {\bibinfo {volume} {75}},\ \bibinfo {pages} {062310}
  (\bibinfo {year} {2007})}\BibitemShut {NoStop}%
\bibitem [{\citenamefont {Yang}\ \emph {et~al.}(2011)\citenamefont {Yang},
  \citenamefont {Wang},\ and\ \citenamefont {Liu}}]{YangFP2011}%
  \BibitemOpen
  \bibfield  {author} {\bibinfo {author} {\bibfnamefont {W.}~\bibnamefont
  {Yang}}, \bibinfo {author} {\bibfnamefont {Z.-Y.}\ \bibnamefont {Wang}}, \
  and\ \bibinfo {author} {\bibfnamefont {R.-B.}\ \bibnamefont {Liu}},\
  }\href@noop {} {\bibfield  {journal} {\bibinfo  {journal} {Front. Phys.}\
  }\textbf {\bibinfo {volume} {6}},\ \bibinfo {pages} {2} (\bibinfo {year}
  {2011})}\BibitemShut {NoStop}%
\bibitem [{\citenamefont {Robledo}\ \emph {et~al.}(2011)\citenamefont
  {Robledo}, \citenamefont {Childress}, \citenamefont {Bernien}, \citenamefont
  {Hensen}, \citenamefont {Alkemade},\ and\ \citenamefont
  {Hanson}}]{RobledoNature2011}%
  \BibitemOpen
  \bibfield  {author} {\bibinfo {author} {\bibfnamefont {L.}~\bibnamefont
  {Robledo}}, \bibinfo {author} {\bibfnamefont {L.}~\bibnamefont {Childress}},
  \bibinfo {author} {\bibfnamefont {H.}~\bibnamefont {Bernien}}, \bibinfo
  {author} {\bibfnamefont {B.}~\bibnamefont {Hensen}}, \bibinfo {author}
  {\bibfnamefont {P.~F.~A.}\ \bibnamefont {Alkemade}}, \ and\ \bibinfo {author}
  {\bibfnamefont {R.}~\bibnamefont {Hanson}},\ }\href@noop {} {\bibfield
  {journal} {\bibinfo  {journal} {Nature}\ }\textbf {\bibinfo {volume} {477}},\
  \bibinfo {pages} {574} (\bibinfo {year} {2011})}\BibitemShut {NoStop}%
\bibitem [{\citenamefont {Pfaff}\ \emph {et~al.}(2013)\citenamefont {Pfaff},
  \citenamefont {Taminiau}, \citenamefont {Robledo}, \citenamefont {Bernien},
  \citenamefont {Markham}, \citenamefont {Twitchen},\ and\ \citenamefont
  {Hanson}}]{PfaffNatPhys2013}%
  \BibitemOpen
  \bibfield  {author} {\bibinfo {author} {\bibfnamefont {W.}~\bibnamefont
  {Pfaff}}, \bibinfo {author} {\bibfnamefont {T.~H.}\ \bibnamefont {Taminiau}},
  \bibinfo {author} {\bibfnamefont {L.}~\bibnamefont {Robledo}}, \bibinfo
  {author} {\bibfnamefont {H.}~\bibnamefont {Bernien}}, \bibinfo {author}
  {\bibfnamefont {M.}~\bibnamefont {Markham}}, \bibinfo {author} {\bibfnamefont
  {D.~J.}\ \bibnamefont {Twitchen}}, \ and\ \bibinfo {author} {\bibfnamefont
  {R.}~\bibnamefont {Hanson}},\ }\href@noop {} {\bibfield  {journal} {\bibinfo
  {journal} {Nat. Phys.}\ }\textbf {\bibinfo {volume} {9}},\ \bibinfo {pages}
  {29} (\bibinfo {year} {2013})}\BibitemShut {NoStop}%
\bibitem [{\citenamefont {Cramer}\ \emph {et~al.}(2016)\citenamefont {Cramer},
  \citenamefont {Kalb}, \citenamefont {Rol}, \citenamefont {Hensen},
  \citenamefont {Blok}, \citenamefont {Markham}, \citenamefont {Twitchen},
  \citenamefont {Hanson},\ and\ \citenamefont {Taminiau}}]{CramerNC2016}%
  \BibitemOpen
  \bibfield  {author} {\bibinfo {author} {\bibfnamefont {J.}~\bibnamefont
  {Cramer}}, \bibinfo {author} {\bibfnamefont {N.}~\bibnamefont {Kalb}},
  \bibinfo {author} {\bibfnamefont {M.~A.}\ \bibnamefont {Rol}}, \bibinfo
  {author} {\bibfnamefont {B.}~\bibnamefont {Hensen}}, \bibinfo {author}
  {\bibfnamefont {M.~S.}\ \bibnamefont {Blok}}, \bibinfo {author}
  {\bibfnamefont {M.}~\bibnamefont {Markham}}, \bibinfo {author} {\bibfnamefont
  {D.~J.}\ \bibnamefont {Twitchen}}, \bibinfo {author} {\bibfnamefont
  {R.}~\bibnamefont {Hanson}}, \ and\ \bibinfo {author} {\bibfnamefont {T.~H.}\
  \bibnamefont {Taminiau}},\ }\href@noop {} {\bibfield  {journal} {\bibinfo
  {journal} {Nat. Commun.}\ }\textbf {\bibinfo {volume} {7}},\ \bibinfo {pages}
  {11526} (\bibinfo {year} {2016})}\BibitemShut {NoStop}%
\bibitem [{\citenamefont {Kalb}\ \emph {et~al.}(2016)\citenamefont {Kalb},
  \citenamefont {Cramer}, \citenamefont {Twitchen}, \citenamefont {Markham},
  \citenamefont {Hanson},\ and\ \citenamefont {Taminiau}}]{KalbNC2016}%
  \BibitemOpen
  \bibfield  {author} {\bibinfo {author} {\bibfnamefont {N.}~\bibnamefont
  {Kalb}}, \bibinfo {author} {\bibfnamefont {J.}~\bibnamefont {Cramer}},
  \bibinfo {author} {\bibfnamefont {D.~J.}\ \bibnamefont {Twitchen}}, \bibinfo
  {author} {\bibfnamefont {M.}~\bibnamefont {Markham}}, \bibinfo {author}
  {\bibfnamefont {R.}~\bibnamefont {Hanson}}, \ and\ \bibinfo {author}
  {\bibfnamefont {T.~H.}\ \bibnamefont {Taminiau}},\ }\href@noop {} {\bibfield
  {journal} {\bibinfo  {journal} {Nat. Commun.}\ }\textbf {\bibinfo {volume}
  {7}},\ \bibinfo {pages} {13111} (\bibinfo {year} {2016})}\BibitemShut
  {NoStop}%
\bibitem [{\citenamefont {Reiserer}\ \emph {et~al.}(2016)\citenamefont
  {Reiserer}, \citenamefont {Kalb}, \citenamefont {Blok}, \citenamefont {van
  Bemmelen}, \citenamefont {Taminiau}, \citenamefont {Hanson}, \citenamefont
  {Twitchen},\ and\ \citenamefont {Markham}}]{ReisererPRX2016}%
  \BibitemOpen
  \bibfield  {author} {\bibinfo {author} {\bibfnamefont {A.}~\bibnamefont
  {Reiserer}}, \bibinfo {author} {\bibfnamefont {N.}~\bibnamefont {Kalb}},
  \bibinfo {author} {\bibfnamefont {M.~S.}\ \bibnamefont {Blok}}, \bibinfo
  {author} {\bibfnamefont {K.~J.~M.}\ \bibnamefont {van Bemmelen}}, \bibinfo
  {author} {\bibfnamefont {T.~H.}\ \bibnamefont {Taminiau}}, \bibinfo {author}
  {\bibfnamefont {R.}~\bibnamefont {Hanson}}, \bibinfo {author} {\bibfnamefont
  {D.~J.}\ \bibnamefont {Twitchen}}, \ and\ \bibinfo {author} {\bibfnamefont
  {M.}~\bibnamefont {Markham}},\ }\href@noop {} {\bibfield  {journal} {\bibinfo
   {journal} {Phys. Rev. X}\ }\textbf {\bibinfo {volume} {6}},\ \bibinfo
  {pages} {021040} (\bibinfo {year} {2016})}\BibitemShut {NoStop}%
\bibitem [{\citenamefont {Ma}\ \emph {et~al.}(2022)\citenamefont {Ma},
  \citenamefont {Li},\ and\ \citenamefont {Liu}}]{MaarXiv2022}%
  \BibitemOpen
  \bibfield  {author} {\bibinfo {author} {\bibfnamefont {W.-L.}\ \bibnamefont
  {Ma}}, \bibinfo {author} {\bibfnamefont {S.-S.}\ \bibnamefont {Li}}, \ and\
  \bibinfo {author} {\bibfnamefont {R.-B.}\ \bibnamefont {Liu}},\ }\href@noop
  {} {\bibfield  {journal} {\bibinfo  {journal} {arXiv:2208.08141v1}\ }
  (\bibinfo {year} {2022})}\BibitemShut {NoStop}%
\end{thebibliography}

\end{document}